\documentclass[reprint,amsmath,amssymb,pre]{revtex4-2}
\usepackage{graphicx}
\usepackage{color}
\usepackage{dcolumn}
\usepackage{bm}
\usepackage{nicefrac}
\usepackage{hyperref}
\usepackage[textsize=tiny]{todonotes}

\renewcommand{\eqref}[1]{Eq.~(\ref{#1})}
\newcommand{\figref}[1]{Fig.~\ref{#1}}
\newcommand{\appref}[1]{Appendix~\ref{#1}}

\begin{document}
\title{Dynamical phase transition in the growth of programmable polymorphic materials}

\author{Fan Chen}
\affiliation{Department of Chemistry, Princeton University, Princeton, NJ 08544, USA}
\author{William M.~Jacobs}
\email{wjacobs@princeton.edu}
\affiliation{Department of Chemistry, Princeton University, Princeton, NJ 08544, USA}

\date{\today}

\begin{abstract}
  The hallmark feature of polymorphic systems is their ability to assemble into many possible structures at the same thermodynamic state.
  Designer polymorphic materials can in principle be engineered via programmable self-assembly, but the robustness of the assembly process depends on dynamical factors that are poorly understood.
  Here we predict a new failure mode for the growth of multicomponent polymorphic materials, in which dynamical coexistence occurs between ordered and disordered assembly trajectories.
  We show that this transition is preceded by the formation of a steady-state disordered wetting layer, suggesting a nonequilibrium analogy to pre-melting phenomena at equilibrium.
  This dynamical phase transition is likely to occur in a variety of systems and may fundamentally limit the complexity of polymorphic materials that can be designed through programmable self-assembly.
\end{abstract}

\maketitle

\section{Introduction}

Polymorphic and multiphasic materials, such as organic crystals~\cite{nangia2008conformational}, colloidal crystals~\cite{jacobs2025assembly}, and biomolecular mixtures in living cells~\cite{banani2017biomolecular}, are characterized by the possibility of assembling many distinct structures under the same thermodynamic conditions.
\textit{Programmable polymorphic materials}~\cite{murugan2015multifarious,sartori2020assembly,jacobs2021self} are an emerging class of complex materials in which multiple distinct polymorphs are encoded by design, for example via sequence-programmable interactions among polymeric subunits~\cite{evans2024pattern,chen2024emergence}.
These systems generically exhibit a storage capacity, defined as the maximum number of distinct phases that can be encoded via the programmable interactions among the various subunits.
This limit arises from the finite number of degrees of freedom that are available to be tuned to ensure that each of the target polymorphs is thermodynamically stable~\cite{murugan2015multifarious,sartori2020assembly,jacobs2021self}.

Yet in practice, materials must be assembled under nonequilibrium conditions~\cite{whitelam2015pathway,sanz2007evidence,jacobs2016self}.
To promote the assembly of a particular programmed polymorph, the conditions must be chosen such that a seeded nucleation and growth pathway is kinetically accessible, while the nucleation of alternative polymorphs is suppressed.
For finite-size target structures, controlling the nucleation step can be sufficient~\cite{murugan2015multifarious,sartori2020assembly,evans2024pattern}.
However, assembling \textit{spatially unlimited} polymorphic materials also requires control over the steady-state growth dynamics, which can fail in multiple ways.
Nucleation of polymorphs that differ from the seed can occur in the bulk, as is commonly observed in colloidal crystal growth~\cite{Hensley2022self,Hensley2023macroscopic,landy2023programming}, or at the interface of the growing seeded material.
Increasing the supersaturation to drive faster growth is also typically expected to increase the rate of defect incorporation~\cite{whitelam2010control,whitelam2012multicomponent}, which can ultimately lead to the formation of microstructures that differ qualitatively from the equilibrium polymorphs~\cite{whitelam2014critical,nguyen2016design}.
Thus in general, understanding the growth dynamics of spatially unlimited materials requires us to look beyond the near-equilibrium self-assembly paradigm.

Here we predict a qualitatively different failure mode, in which the seeded growth of a multicomponent polymorphic material undergoes a first-order \textit{dynamical} phase transition between ordered and disordered assembly trajectories.
We first introduce a minimal model to describe the steady-state growth of polymorphic materials.
This model maps out the necessary conditions for stable seeded growth, including a parameter regime in which seeded and disordered polymorphs can potentially grow at the same velocity.
We then verify this prediction by simulating a lattice model and directly observing dynamical coexistence during steady-state growth when the number of encoded polymorphs is close to the storage capacity.
Similarly to equilibrium solid--solid transitions~\cite{selke1983potts,wang2023polymorphic}, the approach to this dynamical transition features the formation of a nonequilibrium wetting layer at the interface of the growing structure.
The generality of our model suggests that this failure mode may constrain seeded growth in a variety of polymorphic systems poised near capacity, establishing an important design rule for realizing programmable polymorphic materials in practice.

\section{Dynamical mean-field model}

\begin{figure*}
  \includegraphics[width=2\columnwidth]{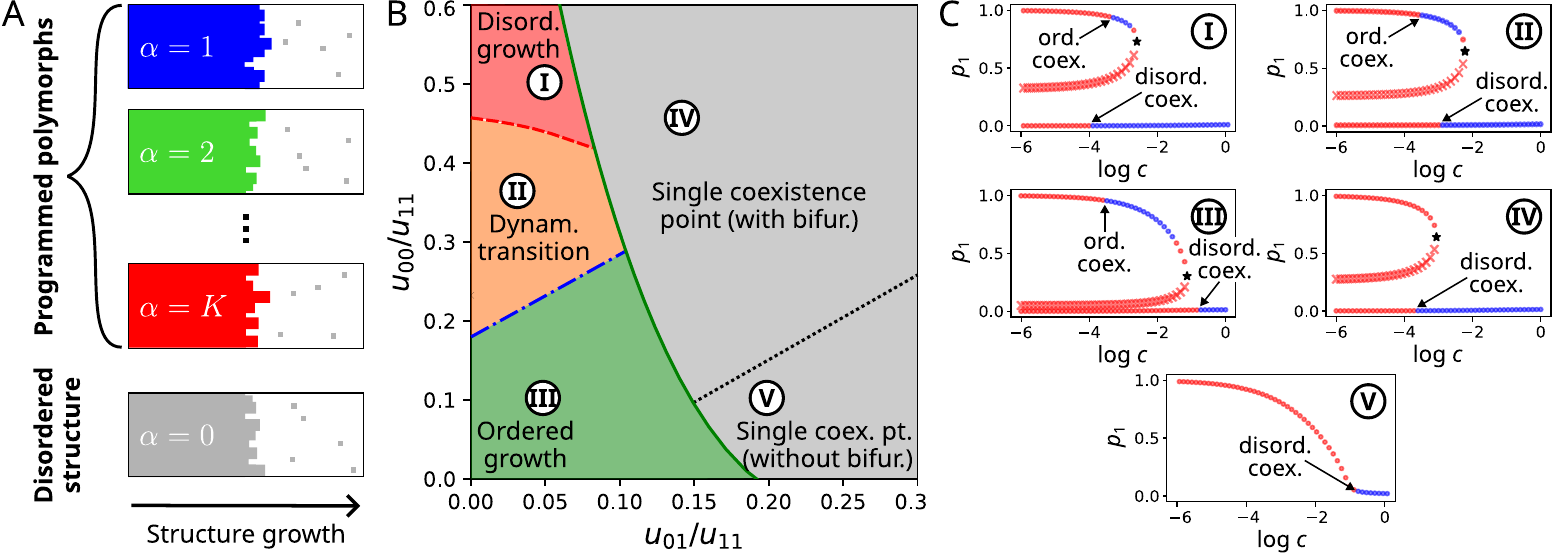}
  \caption{\textbf{A dynamical mean-field model predicts crystal growth behaviors for seeded polymorphic self-assembly.}
    (A)~Seeded growth of programmed polymorphs $\alpha=1,\ldots,K$ and a disordered structure, $\alpha = 0$, in contact with a dilute mixture.
    (B)~Dynamical phase diagram showing different growth regimes with $u_{11} = -8$, $\gamma_1 = 1/64$, and a single ordered structure ($K = 1$).
    Each region of the diagram corresponds to the qualitative assembly behavior predicted by the relationships among the ordered coexistence, disordered coexistence, and bifurcation points in a steady-state bifurcation diagram.
    See the text for descriptions of the boundaries between these regions.
    (C)~Representative steady-state bifurcation diagrams for each of the growth regimes, corresponding to the labeled points in panel B.
    Circles and crosses indicate stable and unstable branches, respectively, of the solution to $\dot p_1 = 0$.
    Blue points indicate concentrations at which the growth rate is positive ($\dot N > 0$), whereas red points indicate where the growth rate is negative ($\dot N < 0$).
    Coexistence points ($\dot N = 0$) are shown for both the ordered (ord) and disordered (disord) branches.
    The bifurcation point is indicated by a black star.
    \textit{Case I:} Disordered coexistence occurs at a lower concentration than ordered coexistence.  Because the disordered growth velocity increases rapidly with concentration, dynamical coexistence is generally not possible.
    \textit{Case II:} The ordered coexistence point occurs at a lower concentration than the disordered coexistence point, leading to the possibility of dynamical coexistence at a concentration below the bifurcation point.
    \textit{Case III:} Both the ordered coexistence point and the bifurcation point occur at lower concentrations than the disordered coexistence point.
    \textit{Case IV:} Only one disordered coexistence point exists. Consequently, only disordered growth is possible.
    \textit{Case V:} No bifurcation point exists.
  \label{fig:model}}
\end{figure*}

\subsection{Model of polymorphic crystal growth}

We first consider a minimal model of programmable polymorphic growth.
Encoded polymorphs may correspond to multicomponent crystals with different subunit permutations, unit cell geometries, compositions, or some combination thereof.
By construction, each of the $K$ encoded polymorphs can be distinguished by its local composition and/or configurational order.
To understand the conditions required for selective self-assembly of a particular encoded polymorph, we study the growth of an initial seed of a target polymorph in contact with a dilute phase with fixed subunit concentrations. 
The seed polymorph grows when the dilute phase is supersaturated, so that freely diffusing and rotating subunits in the dilute phase are incorporated into the material via an interface that advances at constant average velocity (\figref{fig:model}A)~\footnote{This assumption of constant supersaturation, resulting in steady-state growth, can be straightforwardly achieved in experiments by monitoring growth in real time~\cite{Hensley2023macroscopic}.}.
However, the structure that assembles via this process may not match the initial seed, since subunits can be incorporated into the growing structure in incorrect positions or orientations relative to the seed polymorph.
In what follows, we will be interested in the steady-state growth behavior of this assembly process.

The key simplifying assumption of our analysis is that each subunit that is added at the interface can be uniquely assigned either to one of the $K$ encoded ``ordered'' polymorphs, or to none of them, based on its local environment within the growing condensed phase.
This means that the rate of subunit addition to polymorph $\alpha$ via the arrival of subunits at the interface is $k_{\text{attach}} \propto \gamma_\alpha c$, where $c$ is the total subunit concentration in the dilute phase and $\gamma_\alpha$ is the probability that a random orientation of a randomly chosen subunit is consistent with polymorph $\alpha$.
This growth process is reversible in the sense that a subunit can detach from the interface and diffuse back to the dilute phase.
Moreover, this assembly process is reaction-limited, since diffusion in the constant-concentration dilute phase is assumed to be fast compared to subunit attachment and detachment, meaning that many attempts are typically required before a subunit is incorporated into the growing structure under the conditions of interest.
Detachment requires breaking intersubunit bonds at the interface and is thus assumed to follow first-order kinetics with rates of the form $k_{\text{detach}} \propto \exp(-\Delta E/k_{\text{B}}T)$, where $\Delta E$ is the increase in energy upon subunit detachment and $k_{\text{B}}T$ is the reduced temperature~\footnote{The qualitative behavior is insensitive to the explicit inclusion of surface diffusion at the solid--dilute interface.  See \textit{Supplementary Information} for further details.}.

We incorporate these subunit attachment and detachment rates into a dynamical mean field model~\cite{whitelam2014critical} that accounts for the local composition and/or configurational order at the interface.
Specifically, the local environment at the interface is described by a polymorph probability vector $\vec p = (p_0, p_1, \ldots, p_K)$ representing the probability that a randomly selected subunit is locally aligned with its neighbors in accordance with any one of the polymorphs $\alpha = 0, \ldots, K$.
Polymorph 0 represents a ``disordered'' state, encompassing all local configurations that do not correspond to any of the encoded polymorphs.
The net growth velocity of each polymorph is thus
\begin{equation}
    \Gamma_\alpha =    
    \textstyle \gamma_\alpha c - p_\alpha\exp\left(\sum_{\beta=0}^K u_{\alpha\beta}p_\beta\right),
    \quad \alpha = 0, \ldots, K.
    \label{eq:dynamical-mf}
\end{equation}
The first term in \eqref{eq:dynamical-mf} accounts for subunit attachment from the dilute phase, where $\gamma_0 = 1 - \sum_{\beta=1}^K \gamma_\beta$ by conservation.
The second term in \eqref{eq:dynamical-mf} accounts for subunit detachment from the interface, assuming a mean-field form for the subunit interaction energies given a local environment described by the polymorph probability vector $\vec p$.
The matrix $\bm{u}$ defines dimensionless coupling coefficients among the encoded and disordered polymorphs, which we describe below.
Finally, the net influx of subunits to the interface must be balanced by the outflux of subunits into the bulk material as the structure grows.
The time evolution of the polymorph probability vector at the interface is therefore given by
\begin{equation}
  \dot{p_\alpha} = \Gamma_\alpha - p_\alpha \dot{N},
  \quad \alpha = 0, \ldots, K,
  \label{eq:mf-eqs}
\end{equation} 
where $\dot{N} = \sum_{\beta=0}^K \Gamma_\beta$ is the growth velocity.
At steady state, the polymorph probability vector at the growth front does not change in time, so that $\dot {\vec p} = 0$.
Depending on the coupling coefficients $\bm{u}$ and the dilute-phase concentration $c$, the growth velocity may be positive, negative, or zero at steady state, as we discuss next.

\subsection{Dynamical phase diagram for one ordered polymorph}

We first consider a single ordered state, for which $K = 1$, and later discuss how the results of this model can be used to understand polymorphic crystal growth more generally.
In the $K = 1$ case, the system of nonlinear equations, \eqref{eq:mf-eqs}, reduces to a single master equation for the probability of finding the interface in the ordered state, $p_1$, at steady state.
Three parameters then characterize the assembly dynamics: the order--order interaction, $u_{11}$; the disorder--disorder interaction, $u_{00}$; and the ``cross-talk'' interaction, $u_{01}$.
In what follows, we describe the construction of a \textit{dynamical phase diagram} (\figref{fig:model}B) that characterizes the assembly behavior as a function of these interaction parameters.
Each region of this diagram describes the qualitative features of the assembly behavior predicted by a bifurcation diagram (\figref{fig:model}C), which depicts the steady-state solution to \eqref{eq:mf-eqs} as a function of the dilute-phase concentration, $c$.

We first consider steady-state solutions at coexistence, meaning that the growth velocity is zero ($\dot N = 0$) and the entire system is at equilibrium.
Under this constraint, the steady-state solution of the $K=1$ master equation takes the form of a generalized Lambert function~\cite{Mezo2016,Mezo2017lambert}, which can have multiple branches of solutions (see \appref{app:mf-model}).
Given a fixed order--order interaction, a one-dimensional curve (solid line in \figref{fig:model}B) indicates where these branches merge, so that the number of stable coexistence points goes from two (colored regions) to one (gray regions).
The former scenario is of interest because it implies that both the ordered and disordered structures can be in equilibrium with the dilute phase, although these two coexistence points typically occur at different concentrations, $c^{\text{ord}}_{\text{coex}}$ (${\dot N = 0}, {p_1 \simeq 1}$) and $c^{\text{disord}}_{\text{coex}}$ (${\dot N = 0}, {p_1 \simeq 0}$), for a given set of interaction parameters.
Equilibrium three-phase coexistence among the dilute, ordered, and disordered structures occurs when $c^{\text{ord}}_{\text{coex}} = c^{\text{disord}}_{\text{coex}}$ (red dashed line in \figref{fig:model}B).

Increasing or decreasing the dilute-phase concentration relative to a coexistence point drives growth (${\dot N > 0}$) or dissolution (${\dot N < 0}$) of that structure.
These concentration-dependent dynamics are described by the stable branches of a bifurcation diagram, which depicts the steady-state solution ($\dot{\vec p} = 0$) at a fixed set of interaction parameters (\figref{fig:model}C).
Whenever there are two coexistence points, there is a local bifurcation of the saddle-point type that separates the stable ordered (${p_1 \simeq 1}$) and disordered branches (${p_1 \simeq 0}$) at steady state (cases I, II, and III in \figref{fig:model}C).
The bifurcation point occurs at a concentration $c_{\text{bifur}}$.
Note that a bifurcation point can exist even when there is only one coexistence point; this scenario occurs if the growth velocity is negative at all dilute-phase concentrations on the stable ordered branch of the bifurcation diagram (case IV in \figref{fig:model}C).

The most likely assembly trajectory corresponds to the polymorph with the highest steady-state growth velocity at a particular dilute-phase concentration.
Focusing on the colored region in \figref{fig:model}B, we classify the qualitative bifurcation diagrams according to the relationships among $c^{\text{ord}}_{\text{coex}}$, $c^{\text{disord}}_{\text{coex}}$, and $c_{\text{bifur}}$ (see \appref{app:phase-diagram}).
If $c^{\text{disord}}_{\text{coex}} < c^{\text{ord}}_{\text{coex}}$, then the disordered structure almost always grows faster for $c > c^{\text{ord}}_{\text{coex}}$ because $\partial \dot N / \partial c$ is much larger for disordered growth (red region in \figref{fig:model}B and case I in \figref{fig:model}C).
We therefore predict that only disordered growth will be observed under these conditions.
However, if $c^{\text{ord}}_{\text{coex}} < c_{\text{bifur}} < c^{\text{disord}}_{\text{coex}}$, then only the ordered structure can grow at concentrations below the bifurcation point (green region in \figref{fig:model}B and case III in \figref{fig:model}C).
In this case, we predict that the ordered structure will grow without competition from the disordered structure for all concentrations between $c^{\text{ord}}_{\text{coex}}$ and $c_{\text{bifur}}$.

\subsection{Dynamical phase transition between ordered and disordered crystal growth}

\begin{figure}
  \includegraphics[width=0.95\columnwidth]{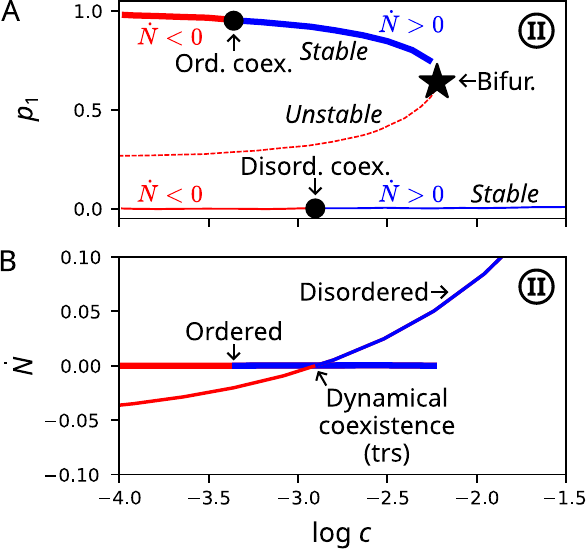}
  \caption{\textbf{Prediction of a dynamical first-order phase transition between ordered and disordered crystal growth.}
    (A)~Example steady-state bifurcation diagram within the dynamical transition region, reproduced from \figref{fig:model}C,II.
    The stable ordered ($p_1 \simeq 1$) and disordered ($p_1 \simeq 0$) branches are colored according to whether the structure grows (blue, $\dot N > 0$) or dissolves (red, $\dot N < 0$) as a function of the dilute-phase concentration, $c$.
    The ordered branch terminates at the bifurcation point.
    (B)~The growth velocity as a function of the dilute-phase concentration, $c$, showing where dynamical coexistence occurs between ordered and disordered assembly with equal growth velocities.
    Note that the thick solid line representing the steady-state growth velocity of the ordered branch is a slowly yet monotonically increasing function of $c$ that terminates at the bifurcation point.
  \label{fig:dynam-trans}}
\end{figure}

The key prediction of this work is the possibility of observing \textit{dynamical coexistence} between ordered and disordered assembly trajectories for certain interaction parameters.
Specifically, if $c^{\text{ord}}_{\text{coex}} < c^{\text{disord}}_{\text{coex}} < c_{\text{bifur}}$ (orange region in \figref{fig:model}B and case II in \figref{fig:model}C), then the ordered and disordered structures can grow at the same velocity at some concentration $c_{\text{trs}}$.
This scenario is illustrated in \figref{fig:dynam-trans}, where the growth velocity, $\dot N$, is shown for each stable branch as a function of the dilute-phase concentration.
Importantly, these two curves intersect at a positive growth velocity, implying that the fastest growing structure will switch from the ordered to the disordered polymorph as $c$ is increased.
This intersection, occurring at a concentration $c_{\text{trs}} > c^{\text{disord}}_{\text{coex}}$, therefore represents a dynamical phase transition between ordered $(c < c_{\text{trs}})$ and disordered growth $(c > c_{\text{trs}}$).

This dynamical transition is predicted to be first-order because the fastest growing polymorph changes discontinuously as the dilute-phase concentration is tuned through $c_{\text{trs}}$.
More precisely, the steady-state polymorph probability vector that corresponds to the fastest growing polymorph has a discontinuity at $c_{\text{trs}}$, even though the growth velocity is predicted to increase continuously as the dilute-phase concentration is increased.
We note that this behavior contrasts with that of a second-order dynamical phase transition, in which the properties of the growing structure change continuously as the dilute-phase concentration, and thus the growth velocity, are varied.
For example, the dynamical transition that was observed for a two-component model of self-assembly with Ising interactions in Ref.~\cite{whitelam2014critical} is a second-order dynamical critical point because the average composition of the growing structure is continuous at the transition.

To summarize, given the dimensionless order--order interaction $u_{11}$, cross-talk interaction $u_{01}$, and disorder--disorder interaction $u_{00}$, we can predict the assembly dynamics for the $K=1$ model and represent the qualitative growth behavior using a dynamical phase diagram.
This analysis predicts that a first-order dynamical phase transition can occur within a particular region of dimensionless interaction parameters; however, the dilute-phase concentration must be tuned to a specific $\bm{u}$-dependent value, $c_{\text{trs}}$, in order to observe dynamical coexistence between ordered and disordered assembly trajectories.

\subsection{Dynamical phase diagram with many encoded polymorphs}

When multiple ordered polymorphs are present ($K > 1$), the steady-state dynamics for any particular ordered polymorph remain qualitatively the same as in the $K = 1$ case if the order--order interactions, $u_{\alpha\alpha}$, and cross-talk interactions, $u_{0\alpha}$ and $u_{\alpha\ne\beta}$, are degenerate for $\alpha,\beta>0$ (see \appref{app:mf-model-Kall}).
However, the coupling coefficients in \eqref{eq:dynamical-mf}, which represent the mean-field interactions between polymorphs, are typically $K$-dependent.
Specifically, as the number of encoded polymorphs increases, the magnitudes of the disorder--disorder and cross-talk interactions tend to increase, eventually favoring the growth of a disordered structure.
An explicit example will be considered in the following section.
Thus, we can simplify our analysis of polymorphic growth by studying the $K=1$ dynamical phase diagram with a system-specific $K$-dependent parametrization for the coupling coefficients $u_{00}$ and $u_{01}$.
We can also follow this approach to estimate the storage capacity for a system with multiple encoded polymorphs, which occurs when the $K$-dependent parametrization crosses into a region of the dynamical phase diagram in which ordered growth is not possible.
This calculation is discussed in detail in the \textit{Supplementary Information}.

\section{Simulation results}

\subsection{Lattice model of multicomponent crystal growth}

\begin{figure}
  \includegraphics[width=\columnwidth]{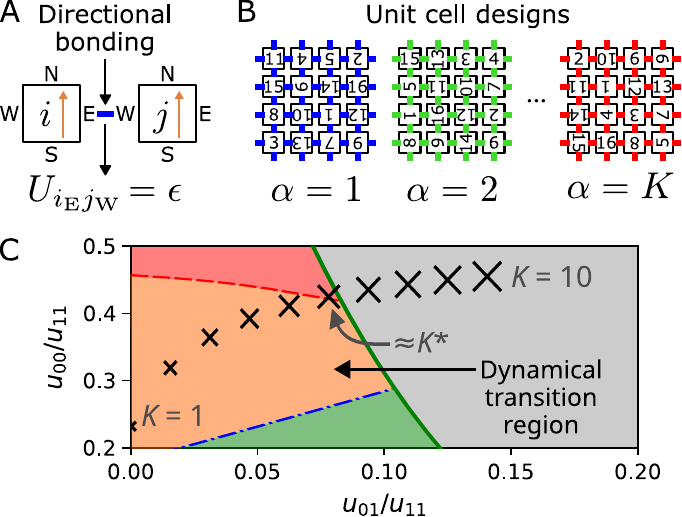}
  \caption{\textbf{Simulating the growth of a multicomponent crystal with directional interactions.} 
    (A)~Tiles associate via specific, directional interactions, which define the nonzero elements of the interaction tensor, $U$.  Here, sticker E on tile $i$ bonds with sticker W on tile $j$.
    (B)~Example unit cell designs for the 16-component system simulated in this work.  The orientation of the each tile is indicated by the rotation of the component number.
    (C)~Mapping between the lattice model and the dynamical mean-field model.  Crosses show the $K$-dependent parametrization of the disorder--disorder interaction, $u_{00}$, and the cross-talk interaction, $u_{01}$, for the lattice model with $1 \le K \le 10$ ordered polymorphs overlaid on the predicted dynamical phase diagram with $u_{11} = -8$ and $\gamma_1 = 1/64$, which is reproduced from \figref{fig:model}B.  The storage capacity occurs at $K^* \simeq 6$, where the parametrized curve crosses the solid green boundary between regions with one or two coexistence points.
  \label{fig:lattice}}
\end{figure}

To verify our predictions, we perform steady-state growth simulations using a minimal model of a multicomponent crystal.
We simulate a two-dimensional lattice model in which ``tiles'' interact through orientation-dependent nearest-neighbor bonds (\figref{fig:lattice}A).
All tiles have $z = 4$ stickers.
We assume that all interactions have the same strength $\epsilon < 0$ and that stickers do not interact nonspecifically.
These rules define a symmetric interaction tensor $U_{i_rj_s}$ in which all elements are either equal to $\epsilon$ or zero, where $i,j \in \{1, \ldots, n\}$ index the $n$ tile types (``components'') and $r,s \in \{\text{N},\text{E},\text{S},\text{W}\}$ index the stickers.
Assuming that each polymorph contains exactly one copy of each component, we encode a unit cell design by arranging and orienting the components to form $zn/2$ unique bonds, including bonds between tiles on the edges of the unit cell and the periodic images of their neighbors~\cite{jacobs2015self} (\figref{fig:lattice}B).
By permuting and rotating the tiles, we can encode multiple unit cell designs, yielding a total interaction tensor $U$ that is the union of the interaction tensors for the unit cell designs, $\{U^{(\alpha)}\}$.
In practice, we generate these designs via simulated annealing to ensure that no shared bonds exist between different unit cells, yielding a total interaction tensor with exactly $Kzn$ non-zero entries, where $K$ is the number of encoded unit cell designs.
This modification of the design strategy introduced in Ref.~\cite{murugan2015multifarious} allows us to encode multiple crystal polymorphs with highly similar thermodynamic stabilities (see \textit{Supplementary Information}).

To simulate the reversible growth of a multicomponent crystal, we seed a $56 \times 1000$ square lattice with a selected polymorph (i.e., a crystal formed from an encoded unit cell design) at one end.
The growing crystal is in contact with a supersaturated dilute phase held at a constant total concentration $c$, where all components have the same concentration $c/n$.
The system is periodic in the transverse dimension, so that a roughly planar interface forms between the crystal and dilute phases.
Assuming that self-assembly is reaction-limited, precluding concentration gradients in the dilute phase~\cite{whitelam2012multicomponent}, we use kinetic Monte Carlo (kMC) simulations~\cite{gillespie1977exact} to model growth at steady state.
Tiles are inserted with random orientations into vacant lattice sites with a rate proportional to the concentration, $k_{\text{insert}} = c/n$.
Following detailed balance~\cite{frenkel2023understanding}, tiles are removed from the lattice with a rate $k_{\text{remove}} = e^{E/k_{\text{B}}T}$, where $E$ is the total energy of all bonds formed by the tile.
Rotations of tiles are implicitly allowed via the removal and reinsertion of the same component with a different orientation.
To model extremely slow diffusion within the bulk crystal, we impose a kinetic constraint in which no state change is allowed for any lattice site whose four neighbors are occupied by tiles~\cite{whitelam2014critical}.
The implementation details and crystalline unit cell designs are described in the \textit{Supplementary Information}.

These simulations can be mapped to the mean-field model by coarse-graining the state of a tile based on its local environment (\figref{fig:lattice}C).
Each tile exists in one of $q = zn$ states, so that $\gamma_\alpha = 1/q$ for $\alpha > 0$.
When crystal growth is ordered, the tile state can be identified as one of the unit cell designs $\alpha = 1, \ldots, K$ based on the bonds it forms with its nearest neighbors.
Assuming that all the ordered crystals have the same stability, the dimensionless bonding energy of a tile that aligns with an ordered state is $u_{\alpha\alpha} \approx z \epsilon / 2k_{\text{B}}T$ for $\alpha > 0$ at the crystal--dilute interface where, on average, each tile forms half as many bonds as it does in the bulk structure.
We approximate the cross-talk interaction as $u_{0\alpha} = u_{\alpha\ne\beta} \approx u_{11} (K-1) / q$ for $\alpha, \beta > 0$ to account for the fact that out of $q$ total tile states, $K-1$ nearest-neighbor states form bonds that are incompatible with the $\alpha$ polymorph.
Finally, we compute the dimensionless energy of a tile within the disordered structure, $u_{00}$, using an equilibrium average over a random local environment, which is appropriate when $c \simeq c^{\text{disord}}_{\text{coex}}$ (see \appref{app:K-dep}).
Importantly, choosing $n = 16$ ($4 \times 4$ unit cells; \figref{fig:lattice}B) and a bond energy $\epsilon = -4k_{\text{B}}T$ places the simulations in the predicted dynamical transition regime.
The resulting $K$-dependent parametrization of the mean-field interactions is shown by cross marks in \figref{fig:lattice}C.
For these parameters, the intersection of the $K$-dependent parametrization curve and the boundary of the dynamical transition region predicts a storage capacity of $K^* \simeq 6$~\footnote{See \textit{Supplementary Information} for an analysis of the mean-field prediction for the storage capacity scaling, which is linear with respect to the number of components in the strong-binding limit when bonds correspond uniquely to programmed polymorphs.}

\subsection{Conditions for seeded self-assembly}

\begin{figure}
  \includegraphics[width=\columnwidth]{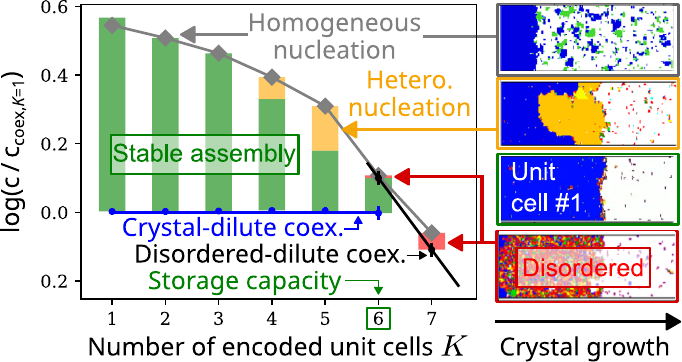}
  \caption{\textbf{Simulated assembly diagram for seeded polymorphic crystal growth.}
    The assembly diagram obtained from kinetic Monte Carlo simulations with $\epsilon = -4 k_{\text{B}}T$ as a function of the number of encoded polymorphs, $K$, and the supersaturation, $c / c_{\text{coex}}$.  The blue and black lines indicate the coexistence conditions for the ordered and disordered structures, respectively.  The gray line indicates the homogeneous nucleation boundary.  On the right, representative snapshots of homogeneous nucleation, heterogeneous nucleation, stable seeded assembly, and disordered assembly are shown from top to bottom.  In each snapshot, bonds are colored according to the encoded unit cells, as shown in \figref{fig:lattice}B.
  \label{fig:diagram}}
\end{figure}

The simulated steady-state growth behavior of the multicomponent crystal model can be summarized by a concentration-dependent assembly diagram (\figref{fig:diagram}).
Using direct-coexistence simulations~\cite{panagiotopoulos2000monte}, we first determine the ordered and disordered coexistence points as a function of $K$, indicating that the storage capacity is indeed $K^* = 6$ for the chosen parameters.
Attempting to encode $K > K^*$ unit cells leads either to dissolution of an ordered crystal or to immediate decomposition into a disordered structure, depending on the dilute-phase concentration.
This behavior is consistent with the transition to the gray region at $K = 7$ in \figref{fig:lattice}C, where only the disordered structure can have a positive growth velocity.

Increasing the supersaturation, ${c / c^{\text{ord}}_{\text{coex}} > 1}$, leads to stable steady-state growth of the ordered seed polymorph for all $K \le K^*$ (\figref{fig:diagram}).
Yet at sufficiently high supersaturations, we observe homogeneous nucleation of crystallites in the dilute phase.
We estimate the boundary beyond which homogeneous nucleation interferes with the growth of the seeded polymorph based on direct observations of nucleation events in our kMC simulations (see \textit{Supplementary Information}).
Due to the strong supersaturation dependence of homogeneous nucleation~\cite{oxtoby1992homogeneous}, decreasing $c$ below this boundary ensures that homogeneous nucleation is an exceedingly rare event.
We also observe heterogeneous nucleation, in which a crystallite that differs from the seed polymorph nucleates at the crystal--dilute interface and grows to span the transverse dimension, leading to a change in the bulk crystal polymorph at the steady-state growth front.
This occurs at supersaturations slightly below the homogeneous nucleation boundary for $K \lesssim K^*$ (see \textit{Supplementary Information}).

However, right at the storage capacity, we observe a \textit{direct transition} from stable seeded self-assembly to disordered structure growth.
This transition differs qualitatively from both heterogeneous nucleation, as crystallites of the encoded polymorphs are rare, and the equilibrium ordered-to-disordered transition between $K=6$ and $K=7$ at coexistence, since it occurs at a positive growth velocity.
Moreover, it differs from the dynamical critical point observed in the growth of two-component structures in Ref.~\cite{whitelam2014critical}.
Instead, the seeded assembly boundary at $K = K^*$ is a first-order dynamical phase transition because the typical growth trajectories change abruptly from ordered to disordered polymorphs as the supersaturation is increased.
This discontinuous jump between qualitatively different assembly trajectories agrees with the key prediction of our mean-field model.

\subsection{Evidence of dynamical coexistence}

\begin{figure}
  \includegraphics[width=\columnwidth]{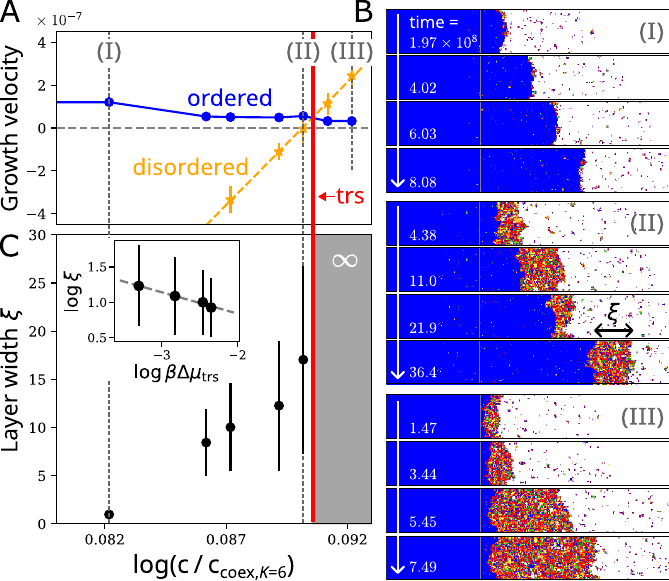}
  \caption{\textbf{A dynamical first-order phase transition limits seeded self-assembly near the storage capacity.}
  (A)~The average growth velocity of ordered (i.e., encoded polymorphs, blue) and disordered (orange) structures in seeded simulations with $K=6$.  The growth velocity of ordered structures is non-monotonic with respect to $c$.  The dynamical transition occurs where the growth velocities intersect at $c_{\text{trs}}$.
  (B)~Example trajectories at points (I) and (II) in the stable seeded assembly regime, showing the disordered wetting layer width, $\xi$, and at point (III) beyond $c_{\text{trs}}$.
  (C)~The average disordered wetting layer width increases with supersaturation below $c_{\text{trs}}$, diverging as $\xi \sim (\Delta\mu_{\text{trs}} / k_{\text{B}}T)^{-0.3}$, where $\Delta\mu_{\text{trs}} / k_{\text{B}}T \equiv \log c_{\text{trs}} - \log c$ (inset).
  \label{fig:transition}}
\end{figure}

Close examination of the assembly behavior at $K = K^*$ reveals bistability in the trajectory ensemble, where it is possible to observe growth of both ordered and disordered structures depending on the initial conditions. 
Specifically, we simulate the $K=6$ mixture within a narrow range of supersaturations and measure the steady-state growth velocities when seeded with either an ordered polymorph or a disordered structure (\figref{fig:transition}A).
Consistent with the mean-field prediction, we find that the growth velocities for ordered and disordered assembly intersect at a positive value, indicating coexistence of ordered and disordered growth trajectories at the dynamical transition point $c_{\text{trs}}$.
Moreover, because the disordered growth velocity increases rapidly with concentration, $c_{\text{trs}}$ occurs close to $c^{\text{disord}}_{\text{coex}}$ as predicted.

Yet unlike in the mean-field model, we find that the growth velocity of the target polymorph decreases as $c_{\text{trs}}$ is approached from below due to the formation of a disordered wetting layer at the crystal--dilute interface.
As the supersaturation is increased, disordered defects at the crystal--dilute interface (\figref{fig:transition}B,I) accumulate into a fluctuating, finite-width wetting layer at steady-state (\figref{fig:transition}B,II).
This feature is also consistent with a first-order transition, as wetting layers commonly appear when approaching phase coexistence in equilibrium systems.
For example, when approaching an equilibrium first-order phase transition between low and high-density phases (or between dilute and condensed phases in solution) from below, a thin layer of the high-density/high-concentration phase can form on a weakly attractive surface under conditions where the high-density/high-concentration phase is thermodynamically unstable~\cite{hansen2013theory}.
In the present scenario, the nonequilibrium disordered phase is dynamically disfavored at $c < c_{\text{trs}}$ because its growth velocity is lower, while the ordered crystal acts as the stabilizing substrate.
Nonetheless, the ordered crystal continues to self-assemble behind this layer, since the dynamical transition occurs at a positive steady-state growth velocity.
The time-averaged thickness of the wetting layer increases as $c$ approaches $c_{\text{trs}}$ (\figref{fig:transition}C) and diverges at the dynamical transition, beyond which disordered assembly dominates (\figref{fig:transition}B,III).
Interestingly, this divergence can be described by a power law close to $c_{\text{trs}}$ (\figref{fig:transition}C,inset), paralleling previous observations of solid--solid wetting layers in equilibrium systems near thermodynamic coexistence points~\cite{selke1983potts,wang2023polymorphic}.

We note that this dynamical phase transition is only observed in simulations at $K=K^*$ with the present interaction energies.
Although this result does not emerge directly from the dynamical mean-field model, it can be understood by noting that the predicted supersaturation required for dynamical coexistence, $c_{\text{trs}} / c^{\text{ord}}_{\text{coex}}$, decreases as $K$ approaches $K^*$.
When fewer polymorphs are encoded, the disordered structure may not attain a positive growth velocity at supersaturations below the conditions where either homogeneous or heterogeneous nucleation occurs in our simulations, preventing the observation of dynamical coexistence at $c_{\text{trs}} > c^{\text{disord}}_{\text{coex}}$.
In particular, in simulations with $4 \le K \le 5$, heterogeneous nucleation at the interface of the growing crystal instead limits the maximum supersaturation for seeded self-assembly, as well as the maximum growth velocity~\footnote{The dependence of the growth velocity on the total concentration is also non-monotonic in the presence of heterogeneous nucleation, but the maximum velocity attained for seeded self-assembly is a monotonic function of $K$.} at which seed polymorphs can be robustly assembled (\figref{fig:diagram}).
Although the heterogeneous nucleation boundary is not predicted directly by steady-state analysis of the mean-field model, these ``switching'' events between bulk crystal polymorphs can be understood as stochastic jumps among the $K$ degenerate stable ordered assembly branches at steady state.

\section{Conclusions}

In conclusion, we present a minimal model of polymorphic growth that predicts a first-order dynamical phase transition between ordered and disordered self-assembly trajectories.
This predicted failure mode is confirmed by simulations of multicomponent crystallization near the storage capacity.
Although some features of our model are qualitatively consistent with prior studies of finite-size structures, in which nucleation controls the assembly outcomes~\cite{murugan2015multifarious,sartori2020assembly,bisker2018retrieval,osat2022multifarious}, the dynamical phase transition that we report only emerges during the steady-state growth of spatially unlimited materials.
Our findings are therefore best conceptualized as a nonequilibrium, multicomponent generalization of equilibrium first-order solid--solid transitions, which are similarly accompanied by diverging wetting layers~\cite{selke1983potts,wang2023polymorphic,li2024polymorphic}.

Our results could inspire the search for analogous dynamical transitions in related studies of polymorphic crystallization~\cite{goodrich2021designing,bupathy2022temperature,chatterjee2024multi} and multiphase droplet growth in multicomponent fluids~\cite{jacobs2021self,chen2023programmable,chen2024emergence}.
Within the present lattice model, different parameter choices could potentially broaden the range of $K$ values over which the transition can be directly observed.
For example, increasing both the number of components and the magnitude of the bond energies is predicted to decrease the spacing between points on the $K$-dependent parametrization, which might reduce the supersaturation required to observe the dynamical transition at multiple values of $K < K^*$ (see \textit{Supplementary Information}).
Furthermore, in the current model and simulations, the specific interaction energies are considered to be the same for all components, and every component has the same multiplicity in each polymorph.
Altering these design parameters could allow for greater control over different regions of the dynamical phase diagram.
However, further study is needed to identify the conditions in which dynamical coexistence is most likely to be observed within a particular model.

Going forward, it will be important to adapt these models to describe specific experimental systems in which polymorphic crystals can be rationally designed through programmable molecular interactions~\cite{jacobs2025assembly}.
In particular, given recent experimental developments in DNA origami~\cite{rothemund2006folding,Sun2020valence}, it should be possible to test these predictions by designing the interactions among micrometer-scale colloidal subunits in multicomponent mixtures~\cite{kahn2022encoding,hayakawa2024symmetry,videbaek2024economical,duque2024limits}.
Studies of this nature will provide unique insights into the relationship between information storage and programmable self-assembly, presenting opportunities to probe the theoretical concepts of cross-talk and storage capacity in ways that are not possible with systems designed to self-assemble into a single ordered structure.
Moreover, studying the failure modes of seeded growth in these systems could provide a deeper understanding of the dynamics of multicomponent crystallization and establish novel routes to engineering polymorphic and reconfigurable materials via programmable self-assembly.

\begin{acknowledgments}
This work is supported by the Princeton Center for Complex Materials (NSF DMR-2011750).
The authors thank W.\ Benjamin Rogers and Mikko Haataja for helpful comments on the manuscript and acknowledge Princeton Research Computing for technical support.
Source code and example calculations can be found in the Supplementary Materials.
\end{acknowledgments}

\clearpage
\appendix

\counterwithin*{figure}{part}
\stepcounter{part}
\renewcommand{\thefigure}{A\arabic{figure}}

\section{$K=1$ dynamical mean-field model}
\label{app:mf-model}

We consider a minimal model in the spirit of Ref.~\cite{whitelam2014critical} in which the assembled structure can be either a single ordered phase ($K=1$) or a disordered phase.
Given $q$ states for each occupied lattice site, the dynamical mean-field equations, \eqref{eq:dynamical-mf} and \eqref{eq:mf-eqs}, become
\begin{subequations}
  \label{eq:mf-eqs-K1}
\begin{align}
  \dot{p_1} &= \Gamma_1 - p_1 (\Gamma_0 + \Gamma_1), \label{eq:dotp}\\
  \Gamma_0 &= \frac{c (q-K)}{q} - (1- p_1) e^{ u_{00} (1-p_1) + u_{01}p_1 }, \label{eq:flux0}\\
\Gamma_1 &= \frac{c}{q} - p_1 e^{ u_{11} p_1 + u_{01}(1-p_1) }.
\label{eq:flux1}
\end{align}
\end{subequations}
The probability of being in the ordered state, $p_1$, at the interface is determined by the equation of motion, \eqref{eq:dotp}, where the net incoming fluxes to the disordered ($\alpha = 0$) and ordered ($\alpha = 1$) structures are given by \eqref{eq:flux0} and \eqref{eq:flux1}, respectively.
The coefficients $u_{00}$, $u_{01}$, and $u_{11}$ are dimensionless interaction parameters, and $c$ is the total dilute-phase concentration in dimensionless units (i.e., relative to the volume occupied by a subunit).
We are interested in obtaining the steady-state solution to \eqref{eq:dotp}, such that $\dot{p_1} = 0$.
As we show in the analysis below, this equation of motion can admit different numbers of solutions, leading to qualitatively different growth behaviors.

\textit{Conditions for coexistence.}
At coexistence, the bulk structure has zero net growth velocity, $\dot{N} \equiv \Gamma_0 + \Gamma_1 = 0$, while also satisfying the steady state criterion $\dot{p_1} = 0$.
From here on, we shall write $p \equiv p_1$ except when this simplified notation would lead to ambiguity.
We can identify the conditions for coexistence by eliminating the concentration, $c$, leading to
\begin{equation}
  1 = \frac{p}{1-p} e^{(u_{11} + u_{00} - 2u_{01})p + u_{01}- u_{00} + \log(q-K)},
  \label{eq:coexEq}
\end{equation}
or, equivalently,
\begin{subequations}
  \begin{align}
    \label{eq:fp}
    f(p) &\equiv p + p e^{a p + b} - 1 = 0, \\
    a &\equiv u_{11} + u_{00} - 2u_{01}, \\
    b &\equiv \log{(q-K)} + u_{01}- u_{00}.
  \end{align}
\end{subequations}
By rewriting \eqref{eq:fp} as
\begin{equation}
p e^{ap} + e^{-b} p = e^{-b}, 
\end{equation}
we see that the coexistence points are given by solutions of the generalized $r$-Lambert function,
\begin{equation}
  \label{eq:Wr}
  p_{\text{coex}} = \frac{1}{a} W_{e^{-b}}(a e^{-b}),
\end{equation}
which may be multivalued.
Since $e^{-b} > 0$, the $r$-Lambert function has either one or three solutions based on the analysis in Ref.~\cite{Mezo2017lambert}.
In particular, there are three solutions to \eqref{eq:Wr} if $ b > 2$ and $g(\beta_r) < a e^{-b} < g(\alpha_r)$, where $ g(x) = x e^{x} + e^{-b}x$, $\alpha_r = W^{(-1)}(-e^{b+1}) - 1$, $\beta_r = W^{(0)} (-e^{b+1}) - 1$, and $W^{(-1)}$/$W^{(0)}$ are the $-1$/$0$ branches of the Lambert $W$ function, respectively.  
Otherwise, there is one solution to \eqref{eq:Wr}.

After solving \eqref{eq:Wr} to obtain one or more values of $p_\text{coex}$, we solve for the concentration at coexistence using \eqref{eq:mf-eqs-K1} with $\Gamma_0 = \Gamma_1 = \dot N = 0$.
When there are three solutions for $p_{\text{coex}}$, two of the solutions correspond to coexistence between the ordered structure and the dilute phase, $p_\text{coex}^{\text{ord}}$, and to coexistence between the disordered structure and the dilute phase, $p_\text{coex}^{\text{disord}}$.
The remaining coexistence solution corresponds to the zero-growth-velocity point on the unstable branch of the bifurcation diagram.
In general, $p_\text{coex}^{\text{ord}}$ and $p_\text{coex}^{\text{disord}}$ occur at different concentrations, $c_\text{coex}^{\text{ord}}$ and $c_\text{coex}^{\text{disord}}$, respectively,
\begin{subequations}
  \label{eq:ccoex}
  \begin{align}
    \log{c_\text{coex}^{\text{ord}}} &= \log{p_\text{coex}^{\text{ord}}} + (u_{11} - u_{01})p_\text{coex}^{\text{ord}} + u_{01} \nonumber \\
    & \quad + \log q, \\
    \log{c_\text{coex}^{\text{disord}}} &= \log (1-p_\text{coex}^{\text{disord}}) + (u_{01} - u_{00})p_\text{coex}^{\text{disord}} \nonumber \\
    & \quad + u_{00} + \log \frac{q}{q-K}.
  \end{align}
\end{subequations}
We can estimate these coexistence concentrations by taking appropriate limits of \eqref{eq:ccoex},
\begin{subequations}
  \begin{align}
    \log c_\text{coex}^{\text{ord}} &\approx \lim_{p \to 1} \log c_\text{coex}^{\text{ord}} = u_{11} + \log q, \label{eq:c_ord}\\
    \log c_\text{coex}^{\text{disord}} &\approx \lim_{p \to 0} \log c_\text{coex}^{\text{disord}} = u_{00}  + \log \frac{q}{q-K}. \label{eq:c_dis}
  \end{align}
\end{subequations}

\textit{Characterization of the bifurcation point.}
When there are two stable coexistence points, there are necessarily two stable branches of steady-state behaviors.
Each branch exhibits a positive growth velocity when $c > c_{\text{coex}}$.
To understand the conditions under which we obtain either stable ordered or disordered growth, and whether we can have dynamical coexistence between ordered and disordered growth, we first need to determine the point at which the stable ordered growth branch terminates.
This point corresponds to the bifurcation point, beyond which only disordered growth is possible.
The bifurcation point occurs at a unique concentration $c_{\text{bifur}}$.

Following the notation in the dynamical systems literature~\cite{Kuznetsov1998}, for a given set of interaction parameters, the equation of motion, \eqref{eq:dotp}, can be written in the form
\begin{equation}
  \dot{p} = g(p, c),
  \label{eq:stability-analysis}
\end{equation}
where the probability of the ordered state $p$ is the independent variable and the concentration $c$ is a control parameter.
The steady-state assembly behaviors correspond to fixed points of this dynamical system.
Within the relevant parameter regions, we find fixed points, $p = p^*$, given a fixed concentration $c = c^*$, by numerically solving $\dot{p} = g(p, c) = 0$.
The bifurcation point occurs where $(\partial g/\partial p)(p^*, c^*) = 0$.
In the vicinity of the bifurcation point, we always find that $(\partial g / \partial c)(p^*, c^*) \neq 0$ and $(\partial^2 g / \partial p^2)(p^*, c^*) \neq 0$. 
Therefore, the bifurcation point is a local bifurcation of the saddle-node type by a straightforward application of Theorem 3.1 in Ref.~\cite{Kuznetsov1998}.

For the minimal $K = 1$ model, the master equation can be written in the form $p = g(p, c)$, where the function $g$ and its derivatives with respect to $p$ and $c$ are
\begin{widetext}
\begin{subequations}
\begin{align}
g(p, c) &= \frac{c}{q} - p \left[ e^{(u_{11} - u_{01})p + u_{01}} - e^{(u_{01} - u_{00})p + u_{00}} + c \right] + p^2 \left[ e^{(u_{11} - u_{01})p + u_{01}} - e^{(u_{01} - u_{00})p + u_{00}} \right], \label{eq:dynamical}\\
\frac{\partial g}{\partial p} &=   (2p - 1) \left[ e^{(u_{11} - u_{01})p + u_{01}} - e^{(u_{01} - u_{00})p + u_{00}} \right] - c  \\
&\quad - p \left[ (u_{11} - u_{01})e^{(u_{11} - u_{01})p + u_{01}} - (u_{01} - u_{00})e^{(u_{01} - u_{00})p + u_{00}} \right] \nonumber  \\ 
&\quad + p^2 \left[ (u_{11} - u_{01})e^{(u_{11} - u_{01})p + u_{01}} - (u_{01} - u_{00})e^{(u_{01} - u_{00})p + u_{00}} \right], \nonumber \\
\frac{\partial g}{\partial c} &= \frac{1}{q} - p.
\end{align}
\end{subequations}
\end{widetext}
In practice, we solve for the bifurcation point numerically by nonlinear least squares.

\section{$K=1$ dynamical phase diagram}
\label{app:phase-diagram}

We next solve for the boundaries between the assembly regimes shown in \figref{fig:model}B.
When two stable coexistence points are present, these regimes are $c^{\text{disord}}_{\text{coex}} < c^{\text{ord}}_{\text{coex}}$ (disordered assembly region), $c^{\text{ord}}_{\text{coex}} < c_{\text{bifur}} < c^{\text{disord}}_{\text{coex}}$ (ordered assembly region), and $c^{\text{ord}}_{\text{coex}} < c^{\text{disord}}_{\text{coex}} < c_{\text{bifur}}$ (dynamical transition region).

\textit{Upper and lower boundaries of the region with both ordered and disordered coexistence.}
Based on the properties of the $r$-Lambert function, the region with three coexistence solutions is enclosed by two curves, such that $a_\text{lower}(b) < a < a_\text{upper}(b)$ [see \eqref{eq:fp}].
The upper and lower bounds are given by
\begin{equation}
  a_\text{upper/lower}(b) = f[W^{(-1)/(0)}(-e^{-b + 1}) - 1, b],
  \label{seq:Wbranch}
\end{equation}
where $f(x, b) \equiv x e^{x+b} + 1$.
These two curves merge where $a^* = -4$ and $b^* = 2$.
\eqref{seq:Wbranch} therefore defines the solid green boundary in \figref{fig:model}B.

\textit{Boundary between the disordered assembly and the dynamical transition regions.}
The disordered assembly region describes the parameter regime for which $c^{\text{disord}}_{\text{coex}} < c^{\text{ord}}_{\text{coex}}$.
Therefore, the boundary between the disordered assembly region and the dynamical transition region is defined by $c^{\text{disord}}_{\text{coex}} = c^{\text{ord}}_{\text{coex}}$.
In the strong-bonding limit, i.e., $\epsilon \rightarrow -\infty$, we can equate the two approximate relations for the coexistence concentrations \eqref{eq:c_ord} and \eqref{eq:c_dis} to yield the approximation
\begin{equation}
  u_{00} \approx \log (q-K) + u_{11}. 
\end{equation}
In \figref{fig:model}B, this boundary, shown by the red dashed curve, is obtained by numerically solving for the steady-state condition $\dot{p} = 0$, the coexistence condition $\dot{N} = 0$, and $c_\text{coex}^{\text{ord}} = c_\text{coex}^{\text{disord}}$ simultaneously.

\textit{Boundary between the ordered assembly and the dynamical transition regions.}
Because the dynamical transition requires $c_\text{coex}^{\text{ord}} < c_\text{coex}^{\text{disord}} < c_\text{bifur}$, the boundary between the ordered assembly region and the dynamical transition region is given by the condition $c_\text{coex}^{\text{disord}} = c_\text{bifur}$.
To locate this boundary numerically at a fixed value of $u_{01}$, we simultaneously solve for $u_{00}$ and $p_{\text{bifur}}$, where
\begin{subequations}
  \begin{align}
    c_{\text{coex}}^{\text{disord}}(u_{00}) &= c_{\text{bifur}}(u_{00}), \label{eq:bondx}\\
    \frac{\partial g}{\partial p}(p_{\text{bifur}}, c_{\text{bifur}}; u_{00}) &= 0,
  \end{align}
\end{subequations}
and $c_\text{bifur}(u_{00})$ is obtained from \eqref{eq:dynamical},
\begin{align}
  c_{\text{bifur}} = \frac{1}{1/q- p_\text{bifur}}  \big[ &p_\text{bifur} ( e^{(u_{11} - u_{01})p_\text{bifur}+ u_{01}} \nonumber \\
    &\qquad - e^{(u_{01} - u_{00})p_\text{bifur} + u_{00}}) \nonumber \\
    &+ p_\text{bifur}^2 ( e^{(u_{11} - u_{01})p_\text{bifur} + u_{01}} \nonumber \\
    &\qquad- e^{(u_{01} - u_{00})p_\text{bifur} + u_{00}} ) \big].
\end{align}
This boundary is shown by the dot-dashed blue line in \figref{fig:model}B.

\section{$K > 1$ dynamical mean-field model}
\label{app:mf-model-Kall}

We now generalize the minimal model to cases where there are $K > 1$ ordered states, all with the same order--order interaction $u_{\alpha\alpha} = u_{11}$ for $\alpha > 0$.
At steady state, we can assume that one of the $K$ ordered states dominates each of the $K$ degenerate ordered stable branches.
By symmetry, the remaining $K-1$ ordered states have equal probabilities on each ordered stable branch.
We therefore introduce a small parameter $s$ such that the non-dominant ordered states have probability $p_\alpha = s/(K-1)$ for ${\alpha = 2, \ldots, K}$, and the dominant ordered state has probability $p_1 =  1 - \sum_{\alpha=2}^K p_\alpha = p - s$.
The probability of the disordered state is then $p_0 = 1 - p$, where $\sum_{\alpha=1}^K p_\alpha = p \leq 1$. 
We further assume that all cross-talk interactions have the same strength, $u_{\alpha \beta} = u_{01}$ for $\alpha \neq \beta$.
This parametrization leads to the following system of dynamical equations [cf.~\eqref{eq:mf-eqs-K1}],
\begin{subequations}
\begin{align}
  \dot{p_1} &= \dot{p} -\dot{s}= \Gamma_1 - (p - s) \dot{N},\\
  \dot{p_\alpha} &= \frac{\dot{s}}{K-1} = \Gamma_2 - \frac{s}{K-1} \dot{N}, \quad \alpha = 2, \ldots, K.
\end{align}
\end{subequations}
Thus, we obtain the steady-state solution by solving
\begin{subequations}
\label{eq:Kdep-dotp}
\begin{align}
  \dot{p} &= -\Gamma_0 + (1 - p) \dot{N} = 0, \\
  \dot{s} &= (K-1)\Gamma_2 - s \dot{N} = 0,
\end{align}
\end{subequations}
where the net incoming fluxes, $\Gamma_0, \Gamma_1, \ldots, \Gamma_K$, and the overall growth rate, $\dot N$, are given by
\begin{subequations}
\begin{align}
\Gamma_0 & = \frac{c}{q}(q-K) - (1-p) e^{ (u_{01}-u_{00}) p + u_{00}}, \\
\Gamma_1 &= \frac{c}{q} - (p-s) e^{ (u_{11}-u_{01})(p - s) + u_{01}} , \\
\Gamma_2 &= \Gamma_3 = \ldots = \Gamma_K \nonumber \\
&= \frac{c}{q} - \frac{s}{K-1} e^{ (u_{11} - u_{01})s / (K-1)}, \\
\dot{N} &= \sum_{\alpha = 0}^{K} \Gamma_\alpha = c - (1-p) e^{ (u_{01}-u_{00}) p + u_{00}} \nonumber \\
 &\qquad\qquad\qquad - (p-s) e^{ (u_{11}-u_{01})(p - s) + u_{01}} \nonumber \\
 &\qquad\qquad\qquad - s e^{ (u_{11} - u_{01})s/(K-1)}.
\end{align}
\end{subequations}

\begin{figure}[t]
    \centering
    \includegraphics[width=\columnwidth]{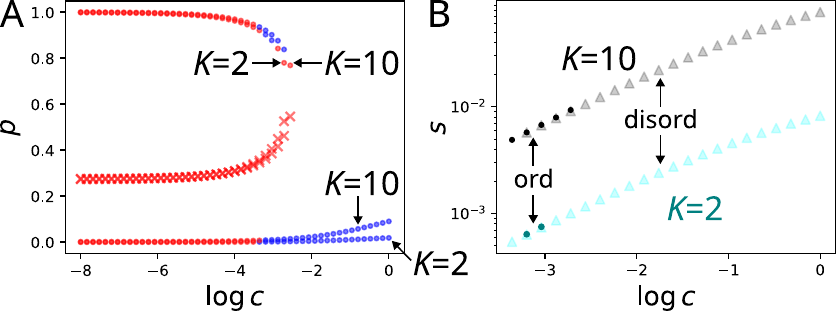}
    \caption{\textbf{Generalization of the $K=1$ mean-field model to $K > 1$.}
      (A)~Numerically determined bifurcation diagrams for $K = 2$ and $K = 10$ using interaction parameters $u_{11} = -8$, $u_{01} = -0.624$, and $u_{00} = -3.4$, which correspond to the parameters near $K^*$ in \figref{fig:lattice}C in the main text.
      (B)~The steady-state value of $s$ along the $K$ degenerate ordered branches (dots) and the disordered branch (triangles) as a function of the dilute-phase concentration $c$. The overlap of dots and triangles corresponds to the region of dynamical coexistence.}
    \label{fig:Kdep-mf}
\end{figure}

In \figref{fig:Kdep-mf}, we directly compute numerical solutions of \eqref{eq:Kdep-dotp} with $u_{11} = -8$ to evaluate the bifurcation diagram as a function of $K > 1$.
Within the dynamical transition region, these diagrams are nearly indistinguishable from the $K=1$ bifurcation diagram (i.e., Fig.~1C in the main text), since the steady-state value of $s$ is small on all stable branches.
This analysis confirms that the $K$-dependence primarily enters the model via the parametrization of the cross-talk and disordered interactions.

\section{Parametrization of $K$-dependent mean-field interactions}
\label{app:K-dep}

The mean-field interaction parameters for the lattice model are motivated by considering the average bonding energy between a tile at a tagged lattice site in coarse-grained state $\alpha$ and each of its $z_{\text{int}}$ nearest neighbors at the solid--dilute phase interface, ignoring higher-order correlations among the nearby lattice sites that constitute the tagged lattice site's local environment.
More precisely, ignoring higher-order correlations means that we can assume that the alignment of a tile with its local environment can be determined by comparing the state of the tile at the tagged lattice site with the state of the tile at a reference nearest-neighbor lattice site.
When the tagged tile is in a coarse-grained state that aligns with the reference tile (i.e., $\alpha = \beta \neq 0$), the bond energy per neighbor is $\epsilon$, and the dimensionless ordered interaction parameter is therefore $u_{\alpha\alpha} = z_{\text{int}}\epsilon / k_{\text{B}}T$.
By contrast, there are $K-1$ possible bonds that can potentially form by chance when the tagged and reference tiles are misaligned.
If the reference tile is assigned to an ordered state $\beta$, then it could be in one of $q-1$ possible states.
The expectation value for the bond energy in this case (i.e., $\alpha \neq \beta \neq 0$) is thus $u_{\alpha\beta} = \epsilon (K - 1) / (q-1) k_{\text{B}}T$.
If the reference tile is instead assigned to the disordered state, meaning that we have no information about its state relative to its neighbors, then it could be in any of the $q$ possible states.
The expectation value for the bond energy in this case (i.e., $\alpha \neq \beta = 0$) is thus $u_{\alpha 0} = \epsilon (K - 1) / q k_{\text{B}}T$.
The dimensionless cross-talk interaction parameters are therefore $u_{\alpha\beta} = u_{\alpha\alpha} (K - 1) / (q-1)$ for $\alpha\ne\beta\ne 0$ and $u_{\alpha 0} = u_{\alpha\alpha} (K - 1) / q$ for $\alpha\ne\beta = 0$.
However, since the difference between these two cross-talk interaction parameters is negligible for large $q$, we use $u_{\alpha\beta} = u_{\alpha 0}$ in the main text for simplicity.

When a tile is in the disordered state, it cannot be aligned with the reference tile by definition.
Therefore, it can form at most $z_{\text{int}} - 1$ bonds within the mean-field approximation.
In principle, the steady-state distribution of the number of bonds, $n_{\text{b}}^{\text{disord}}$, formed by disordered tiles is dependent on the growth rate.
However, because we are most interested in analyzing scenarios when $c \approx c_{\text{coex}}^{\text{disord}}$, at which point the disordered structure is approximately in coexistence with the dilute phase, we can invoke the equilibrium statistical mechanics relation
\begin{equation}
  \langle n_{\text{b}}^{\text{disord}} \rangle \approx \frac{\sum_{n_{\text{b}}=0}^{z_{\text{int}}-1} n_{\text{b}} r(n_{\text{b}}) e^{-n_{\text{b}} \epsilon / k_{\text{B}}T}}{\sum_{n_{\text{b}}=0}^{z_{\text{int}}-1} r(n_{\text{b}}) e^{-n_{\text{b}} \epsilon / k_{\text{B}}T}},
\end{equation}
where the entropic factor $r(n_{\text{b}})$ accounts for the number of ways that $n_{\text{b}}$ bonds could appear out of $z_{\text{int}} - 1$ potentially bonding nearest neighbors in the absence of correlations.
In a two-dimensional system at the interface between phases, $z_{\text{int}} \approx 2$.
Randomly oriented nearest-neighbor tiles form a bond with probability $K/q$, so $r(0) = (q - K)/q$ and $r(1) = K / q$.
The dimensionless disordered--disordered interaction parameter is thus
\begin{equation}
  \label{eq:u00}
  u_{00} \approx \epsilon \langle n_{\text{b}}^{\text{disord}} \rangle / k_{\text{B}}T = \frac{u_{11}}{2} \left[1 + \frac{q-K}{K} e^{u_{11} / 2}\right]^{-1}\!\!\!.
\end{equation}
For large $K$ (i.e., $K \rightarrow q$), $u_{00} / u_{11}$ approaches $\nicefrac{1}{2}$.

\clearpage


\begin{thebibliography}{46}%
\makeatletter
\providecommand \@ifxundefined [1]{%
 \@ifx{#1\undefined}
}%
\providecommand \@ifnum [1]{%
 \ifnum #1\expandafter \@firstoftwo
 \else \expandafter \@secondoftwo
 \fi
}%
\providecommand \@ifx [1]{%
 \ifx #1\expandafter \@firstoftwo
 \else \expandafter \@secondoftwo
 \fi
}%
\providecommand \natexlab [1]{#1}%
\providecommand \enquote  [1]{``#1''}%
\providecommand \bibnamefont  [1]{#1}%
\providecommand \bibfnamefont [1]{#1}%
\providecommand \citenamefont [1]{#1}%
\providecommand \href@noop [0]{\@secondoftwo}%
\providecommand \href [0]{\begingroup \@sanitize@url \@href}%
\providecommand \@href[1]{\@@startlink{#1}\@@href}%
\providecommand \@@href[1]{\endgroup#1\@@endlink}%
\providecommand \@sanitize@url [0]{\catcode `\\12\catcode `\$12\catcode
  `\&12\catcode `\#12\catcode `\^12\catcode `\_12\catcode `\%12\relax}%
\providecommand \@@startlink[1]{}%
\providecommand \@@endlink[0]{}%
\providecommand \url  [0]{\begingroup\@sanitize@url \@url }%
\providecommand \@url [1]{\endgroup\@href {#1}{\urlprefix }}%
\providecommand \urlprefix  [0]{URL }%
\providecommand \Eprint [0]{\href }%
\providecommand \doibase [0]{https://doi.org/}%
\providecommand \selectlanguage [0]{\@gobble}%
\providecommand \bibinfo  [0]{\@secondoftwo}%
\providecommand \bibfield  [0]{\@secondoftwo}%
\providecommand \translation [1]{[#1]}%
\providecommand \BibitemOpen [0]{}%
\providecommand \bibitemStop [0]{}%
\providecommand \bibitemNoStop [0]{.\EOS\space}%
\providecommand \EOS [0]{\spacefactor3000\relax}%
\providecommand \BibitemShut  [1]{\csname bibitem#1\endcsname}%
\let\auto@bib@innerbib\@empty
\bibitem [{\citenamefont {Nangia}(2008)}]{nangia2008conformational}%
  \BibitemOpen
  \bibfield  {author} {\bibinfo {author} {\bibfnamefont {A.}~\bibnamefont
  {Nangia}},\ }\href@noop {} {\bibfield  {journal} {\bibinfo  {journal} {Acc.
  Chem. Res.}\ }\textbf {\bibinfo {volume} {41}},\ \bibinfo {pages} {595}
  (\bibinfo {year} {2008})}\BibitemShut {NoStop}%
\bibitem [{\citenamefont {Jacobs}\ and\ \citenamefont
  {Rogers}(2025)}]{jacobs2025assembly}%
  \BibitemOpen
  \bibfield  {author} {\bibinfo {author} {\bibfnamefont {W.~M.}\ \bibnamefont
  {Jacobs}}\ and\ \bibinfo {author} {\bibfnamefont {W.~B.}\ \bibnamefont
  {Rogers}},\ }\href@noop {} {\bibfield  {journal} {\bibinfo  {journal} {Ann.
  Rev. Cond. Mat. Phys.}\ }\textbf {\bibinfo {volume} {16}},\ \bibinfo {pages}
  {443} (\bibinfo {year} {2025})}\BibitemShut {NoStop}%
\bibitem [{\citenamefont {Banani}\ \emph {et~al.}(2017)\citenamefont {Banani},
  \citenamefont {Lee}, \citenamefont {Hyman},\ and\ \citenamefont
  {Rosen}}]{banani2017biomolecular}%
  \BibitemOpen
  \bibfield  {author} {\bibinfo {author} {\bibfnamefont {S.~F.}\ \bibnamefont
  {Banani}}, \bibinfo {author} {\bibfnamefont {H.~O.}\ \bibnamefont {Lee}},
  \bibinfo {author} {\bibfnamefont {A.~A.}\ \bibnamefont {Hyman}},\ and\
  \bibinfo {author} {\bibfnamefont {M.~K.}\ \bibnamefont {Rosen}},\ }\href@noop
  {} {\bibfield  {journal} {\bibinfo  {journal} {Nat. Rev. Mol. Cell Biol.}\
  }\textbf {\bibinfo {volume} {18}},\ \bibinfo {pages} {285} (\bibinfo {year}
  {2017})}\BibitemShut {NoStop}%
\bibitem [{\citenamefont {Murugan}\ \emph {et~al.}(2015)\citenamefont
  {Murugan}, \citenamefont {Zeravcic}, \citenamefont {Brenner},\ and\
  \citenamefont {Leibler}}]{murugan2015multifarious}%
  \BibitemOpen
  \bibfield  {author} {\bibinfo {author} {\bibfnamefont {A.}~\bibnamefont
  {Murugan}}, \bibinfo {author} {\bibfnamefont {Z.}~\bibnamefont {Zeravcic}},
  \bibinfo {author} {\bibfnamefont {M.~P.}\ \bibnamefont {Brenner}},\ and\
  \bibinfo {author} {\bibfnamefont {S.}~\bibnamefont {Leibler}},\ }\href@noop
  {} {\bibfield  {journal} {\bibinfo  {journal} {Proc. Natl. Acad. Sci.
  U.S.A.}\ }\textbf {\bibinfo {volume} {112}},\ \bibinfo {pages} {54} (\bibinfo
  {year} {2015})}\BibitemShut {NoStop}%
\bibitem [{\citenamefont {Sartori}\ and\ \citenamefont
  {Leibler}(2020)}]{sartori2020assembly}%
  \BibitemOpen
  \bibfield  {author} {\bibinfo {author} {\bibfnamefont {P.}~\bibnamefont
  {Sartori}}\ and\ \bibinfo {author} {\bibfnamefont {S.}~\bibnamefont
  {Leibler}},\ }\href@noop {} {\bibfield  {journal} {\bibinfo  {journal} {Proc.
  Natl. Acad. Sci. U.S.A.}\ }\textbf {\bibinfo {volume} {117}},\ \bibinfo
  {pages} {114} (\bibinfo {year} {2020})}\BibitemShut {NoStop}%
\bibitem [{\citenamefont {Jacobs}(2021)}]{jacobs2021self}%
  \BibitemOpen
  \bibfield  {author} {\bibinfo {author} {\bibfnamefont {W.~M.}\ \bibnamefont
  {Jacobs}},\ }\href@noop {} {\bibfield  {journal} {\bibinfo  {journal} {Phys.
  Rev. Lett.}\ }\textbf {\bibinfo {volume} {126}},\ \bibinfo {pages} {258101}
  (\bibinfo {year} {2021})}\BibitemShut {NoStop}%
\bibitem [{\citenamefont {Evans}\ \emph {et~al.}(2024)\citenamefont {Evans},
  \citenamefont {O’Brien}, \citenamefont {Winfree},\ and\ \citenamefont
  {Murugan}}]{evans2024pattern}%
  \BibitemOpen
  \bibfield  {author} {\bibinfo {author} {\bibfnamefont {C.~G.}\ \bibnamefont
  {Evans}}, \bibinfo {author} {\bibfnamefont {J.}~\bibnamefont {O’Brien}},
  \bibinfo {author} {\bibfnamefont {E.}~\bibnamefont {Winfree}},\ and\ \bibinfo
  {author} {\bibfnamefont {A.}~\bibnamefont {Murugan}},\ }\href@noop {}
  {\bibfield  {journal} {\bibinfo  {journal} {Nature}\ }\textbf {\bibinfo
  {volume} {625}},\ \bibinfo {pages} {500} (\bibinfo {year}
  {2024})}\BibitemShut {NoStop}%
\bibitem [{\citenamefont {Chen}\ and\ \citenamefont
  {Jacobs}(2024)}]{chen2024emergence}%
  \BibitemOpen
  \bibfield  {author} {\bibinfo {author} {\bibfnamefont {F.}~\bibnamefont
  {Chen}}\ and\ \bibinfo {author} {\bibfnamefont {W.~M.}\ \bibnamefont
  {Jacobs}},\ }\href@noop {} {\bibfield  {journal} {\bibinfo  {journal} {J.
  Chem. Theory Comput.}\ }\textbf {\bibinfo {volume} {20}},\ \bibinfo {pages}
  {6881} (\bibinfo {year} {2024})}\BibitemShut {NoStop}%
\bibitem [{\citenamefont {Whitelam}\ and\ \citenamefont
  {Jack}(2015)}]{whitelam2015pathway}%
  \BibitemOpen
  \bibfield  {author} {\bibinfo {author} {\bibfnamefont {S.}~\bibnamefont
  {Whitelam}}\ and\ \bibinfo {author} {\bibfnamefont {R.~L.}\ \bibnamefont
  {Jack}},\ }\href@noop {} {\bibfield  {journal} {\bibinfo  {journal} {Ann.
  Rev. Phys. Chem.}\ }\textbf {\bibinfo {volume} {66}},\ \bibinfo {pages} {143}
  (\bibinfo {year} {2015})}\BibitemShut {NoStop}%
\bibitem [{\citenamefont {Sanz}\ \emph {et~al.}(2007)\citenamefont {Sanz},
  \citenamefont {Valeriani}, \citenamefont {Frenkel},\ and\ \citenamefont
  {Dijkstra}}]{sanz2007evidence}%
  \BibitemOpen
  \bibfield  {author} {\bibinfo {author} {\bibfnamefont {E.}~\bibnamefont
  {Sanz}}, \bibinfo {author} {\bibfnamefont {C.}~\bibnamefont {Valeriani}},
  \bibinfo {author} {\bibfnamefont {D.}~\bibnamefont {Frenkel}},\ and\ \bibinfo
  {author} {\bibfnamefont {M.}~\bibnamefont {Dijkstra}},\ }\href@noop {}
  {\bibfield  {journal} {\bibinfo  {journal} {Phys. Rev. Lett.}\ }\textbf
  {\bibinfo {volume} {99}},\ \bibinfo {pages} {055501} (\bibinfo {year}
  {2007})}\BibitemShut {NoStop}%
\bibitem [{\citenamefont {Jacobs}\ and\ \citenamefont
  {Frenkel}(2016)}]{jacobs2016self}%
  \BibitemOpen
  \bibfield  {author} {\bibinfo {author} {\bibfnamefont {W.~M.}\ \bibnamefont
  {Jacobs}}\ and\ \bibinfo {author} {\bibfnamefont {D.}~\bibnamefont
  {Frenkel}},\ }\href@noop {} {\bibfield  {journal} {\bibinfo  {journal} {J.
  Am. Chem. Soc.}\ }\textbf {\bibinfo {volume} {138}},\ \bibinfo {pages} {2457}
  (\bibinfo {year} {2016})}\BibitemShut {NoStop}%
\bibitem [{\citenamefont {Hensley}\ \emph {et~al.}(2022)\citenamefont
  {Hensley}, \citenamefont {Jacobs},\ and\ \citenamefont
  {Rogers}}]{Hensley2022self}%
  \BibitemOpen
  \bibfield  {author} {\bibinfo {author} {\bibfnamefont {A.}~\bibnamefont
  {Hensley}}, \bibinfo {author} {\bibfnamefont {W.~M.}\ \bibnamefont
  {Jacobs}},\ and\ \bibinfo {author} {\bibfnamefont {W.~B.}\ \bibnamefont
  {Rogers}},\ }\href@noop {} {\bibfield  {journal} {\bibinfo  {journal} {Proc.
  Natl. Acad. Sci. U.S.A.}\ }\textbf {\bibinfo {volume} {119}},\ \bibinfo
  {pages} {e2114050118} (\bibinfo {year} {2022})}\BibitemShut {NoStop}%
\bibitem [{\citenamefont {Hensley}\ \emph {et~al.}(2023)\citenamefont
  {Hensley}, \citenamefont {Videb{\ae}k}, \citenamefont {Seyforth},
  \citenamefont {Jacobs},\ and\ \citenamefont
  {Rogers}}]{Hensley2023macroscopic}%
  \BibitemOpen
  \bibfield  {author} {\bibinfo {author} {\bibfnamefont {A.}~\bibnamefont
  {Hensley}}, \bibinfo {author} {\bibfnamefont {T.~E.}\ \bibnamefont
  {Videb{\ae}k}}, \bibinfo {author} {\bibfnamefont {H.}~\bibnamefont
  {Seyforth}}, \bibinfo {author} {\bibfnamefont {W.~M.}\ \bibnamefont
  {Jacobs}},\ and\ \bibinfo {author} {\bibfnamefont {W.~B.}\ \bibnamefont
  {Rogers}},\ }\href@noop {} {\bibfield  {journal} {\bibinfo  {journal} {Nat.
  Commun.}\ }\textbf {\bibinfo {volume} {14}},\ \bibinfo {pages} {4237}
  (\bibinfo {year} {2023})}\BibitemShut {NoStop}%
\bibitem [{\citenamefont {Landy}\ \emph {et~al.}(2023)\citenamefont {Landy},
  \citenamefont {Gibson}, \citenamefont {Chan}, \citenamefont {Pietryga},
  \citenamefont {Weigand},\ and\ \citenamefont
  {Mirkin}}]{landy2023programming}%
  \BibitemOpen
  \bibfield  {author} {\bibinfo {author} {\bibfnamefont {K.~M.}\ \bibnamefont
  {Landy}}, \bibinfo {author} {\bibfnamefont {K.~J.}\ \bibnamefont {Gibson}},
  \bibinfo {author} {\bibfnamefont {R.~R.}\ \bibnamefont {Chan}}, \bibinfo
  {author} {\bibfnamefont {J.}~\bibnamefont {Pietryga}}, \bibinfo {author}
  {\bibfnamefont {S.}~\bibnamefont {Weigand}},\ and\ \bibinfo {author}
  {\bibfnamefont {C.~A.}\ \bibnamefont {Mirkin}},\ }\href@noop {} {\bibfield
  {journal} {\bibinfo  {journal} {ACS Nano}\ }\textbf {\bibinfo {volume}
  {17}},\ \bibinfo {pages} {6480} (\bibinfo {year} {2023})}\BibitemShut
  {NoStop}%
\bibitem [{\citenamefont {Whitelam}(2010)}]{whitelam2010control}%
  \BibitemOpen
  \bibfield  {author} {\bibinfo {author} {\bibfnamefont {S.}~\bibnamefont
  {Whitelam}},\ }\href@noop {} {\bibfield  {journal} {\bibinfo  {journal}
  {Phys. Rev. Lett.}\ }\textbf {\bibinfo {volume} {105}},\ \bibinfo {pages}
  {088102} (\bibinfo {year} {2010})}\BibitemShut {NoStop}%
\bibitem [{\citenamefont {Whitelam}\ \emph {et~al.}(2012)\citenamefont
  {Whitelam}, \citenamefont {Schulman},\ and\ \citenamefont
  {Hedges}}]{whitelam2012multicomponent}%
  \BibitemOpen
  \bibfield  {author} {\bibinfo {author} {\bibfnamefont {S.}~\bibnamefont
  {Whitelam}}, \bibinfo {author} {\bibfnamefont {R.}~\bibnamefont {Schulman}},\
  and\ \bibinfo {author} {\bibfnamefont {L.}~\bibnamefont {Hedges}},\
  }\href@noop {} {\bibfield  {journal} {\bibinfo  {journal} {Phys. Rev. Lett.}\
  }\textbf {\bibinfo {volume} {109}},\ \bibinfo {pages} {265506} (\bibinfo
  {year} {2012})}\BibitemShut {NoStop}%
\bibitem [{\citenamefont {Whitelam}\ \emph {et~al.}(2014)\citenamefont
  {Whitelam}, \citenamefont {Hedges},\ and\ \citenamefont
  {Schmit}}]{whitelam2014critical}%
  \BibitemOpen
  \bibfield  {author} {\bibinfo {author} {\bibfnamefont {S.}~\bibnamefont
  {Whitelam}}, \bibinfo {author} {\bibfnamefont {L.~O.}\ \bibnamefont
  {Hedges}},\ and\ \bibinfo {author} {\bibfnamefont {J.~D.}\ \bibnamefont
  {Schmit}},\ }\href@noop {} {\bibfield  {journal} {\bibinfo  {journal} {Phys.
  Rev. Lett.}\ }\textbf {\bibinfo {volume} {112}},\ \bibinfo {pages} {155504}
  (\bibinfo {year} {2014})}\BibitemShut {NoStop}%
\bibitem [{\citenamefont {Nguyen}\ and\ \citenamefont
  {Vaikuntanathan}(2016)}]{nguyen2016design}%
  \BibitemOpen
  \bibfield  {author} {\bibinfo {author} {\bibfnamefont {M.}~\bibnamefont
  {Nguyen}}\ and\ \bibinfo {author} {\bibfnamefont {S.}~\bibnamefont
  {Vaikuntanathan}},\ }\href@noop {} {\bibfield  {journal} {\bibinfo  {journal}
  {Proc. Natl. Acad. Sci. U.S.A.}\ }\textbf {\bibinfo {volume} {113}},\
  \bibinfo {pages} {14231} (\bibinfo {year} {2016})}\BibitemShut {NoStop}%
\bibitem [{\citenamefont {Selke}\ and\ \citenamefont
  {Huse}(1983)}]{selke1983potts}%
  \BibitemOpen
  \bibfield  {author} {\bibinfo {author} {\bibfnamefont {W.}~\bibnamefont
  {Selke}}\ and\ \bibinfo {author} {\bibfnamefont {D.~A.}\ \bibnamefont
  {Huse}},\ }\href@noop {} {\bibfield  {journal} {\bibinfo  {journal} {Z. Phys.
  B. Con. Mat.}\ }\textbf {\bibinfo {volume} {50}},\ \bibinfo {pages} {113}
  (\bibinfo {year} {1983})}\BibitemShut {NoStop}%
\bibitem [{\citenamefont {Wang}\ \emph {et~al.}(2023)\citenamefont {Wang},
  \citenamefont {Li}, \citenamefont {Li},\ and\ \citenamefont
  {Han}}]{wang2023polymorphic}%
  \BibitemOpen
  \bibfield  {author} {\bibinfo {author} {\bibfnamefont {X.}~\bibnamefont
  {Wang}}, \bibinfo {author} {\bibfnamefont {B.}~\bibnamefont {Li}}, \bibinfo
  {author} {\bibfnamefont {M.}~\bibnamefont {Li}},\ and\ \bibinfo {author}
  {\bibfnamefont {Y.}~\bibnamefont {Han}},\ }\href@noop {} {\bibfield
  {journal} {\bibinfo  {journal} {Nat. Phys.}\ }\textbf {\bibinfo {volume}
  {19}},\ \bibinfo {pages} {700} (\bibinfo {year} {2023})}\BibitemShut
  {NoStop}%
\bibitem [{Note1()}]{Note1}%
  \BibitemOpen
  \bibinfo {note} {This assumption of constant supersaturation, resulting in
  steady-state growth, can be straightforwardly achieved in experiments by
  monitoring growth in real time~\cite {Hensley2023macroscopic}.}\BibitemShut
  {Stop}%
\bibitem [{Note2()}]{Note2}%
  \BibitemOpen
  \bibinfo {note} {The qualitative behavior is insensitive to the explicit
  inclusion of surface diffusion at the solid--dilute interface. See \protect
  \textit {Supplementary Information} for further details.}\BibitemShut {Stop}%
\bibitem [{\citenamefont {Mez{\H o}}\ and\ \citenamefont
  {Keady}(2016)}]{Mezo2016}%
  \BibitemOpen
  \bibfield  {author} {\bibinfo {author} {\bibfnamefont {I.}~\bibnamefont
  {Mez{\H o}}}\ and\ \bibinfo {author} {\bibfnamefont {G.}~\bibnamefont
  {Keady}},\ }\href@noop {} {\bibfield  {journal} {\bibinfo  {journal} {Eur. J.
  Phys.}\ }\textbf {\bibinfo {volume} {37}},\ \bibinfo {pages} {065802}
  (\bibinfo {year} {2016})}\BibitemShut {NoStop}%
\bibitem [{\citenamefont {Mez{\H o}}\ and\ \citenamefont
  {Baricz}(2017)}]{Mezo2017lambert}%
  \BibitemOpen
  \bibfield  {author} {\bibinfo {author} {\bibfnamefont {I.}~\bibnamefont
  {Mez{\H o}}}\ and\ \bibinfo {author} {\bibfnamefont {{\' A}.}~\bibnamefont
  {Baricz}},\ }\href@noop {} {\bibfield  {journal} {\bibinfo  {journal} {Trans.
  Am. Math. Soc.}\ }\textbf {\bibinfo {volume} {369}},\ \bibinfo {pages} {7917}
  (\bibinfo {year} {2017})}\BibitemShut {NoStop}%
\bibitem [{\citenamefont {Jacobs}\ and\ \citenamefont
  {Frenkel}(2015)}]{jacobs2015self}%
  \BibitemOpen
  \bibfield  {author} {\bibinfo {author} {\bibfnamefont {W.~M.}\ \bibnamefont
  {Jacobs}}\ and\ \bibinfo {author} {\bibfnamefont {D.}~\bibnamefont
  {Frenkel}},\ }\href@noop {} {\bibfield  {journal} {\bibinfo  {journal} {Soft
  Matter}\ }\textbf {\bibinfo {volume} {11}},\ \bibinfo {pages} {8930}
  (\bibinfo {year} {2015})}\BibitemShut {NoStop}%
\bibitem [{\citenamefont {Gillespie}(1977)}]{gillespie1977exact}%
  \BibitemOpen
  \bibfield  {author} {\bibinfo {author} {\bibfnamefont {D.~T.}\ \bibnamefont
  {Gillespie}},\ }\href@noop {} {\bibfield  {journal} {\bibinfo  {journal} {J.
  Phys. Chem.}\ }\textbf {\bibinfo {volume} {81}},\ \bibinfo {pages} {2340}
  (\bibinfo {year} {1977})}\BibitemShut {NoStop}%
\bibitem [{\citenamefont {Frenkel}\ and\ \citenamefont
  {Smit}(2023)}]{frenkel2023understanding}%
  \BibitemOpen
  \bibfield  {author} {\bibinfo {author} {\bibfnamefont {D.}~\bibnamefont
  {Frenkel}}\ and\ \bibinfo {author} {\bibfnamefont {B.}~\bibnamefont {Smit}},\
  }\href@noop {} {\emph {\bibinfo {title} {Understanding molecular simulation:
  from algorithms to applications}}}\ (\bibinfo  {publisher} {Elsevier},\
  \bibinfo {year} {2023})\BibitemShut {NoStop}%
\bibitem [{Note3()}]{Note3}%
  \BibitemOpen
  \bibinfo {note} {See \protect \textit {Supplementary Information} for an
  analysis of the mean-field prediction for the storage capacity scaling, which
  is linear with respect to the number of components in the strong-binding
  limit when bonds correspond uniquely to programmed polymorphs.}\BibitemShut
  {Stop}%
\bibitem [{\citenamefont {Panagiotopoulos}(2000)}]{panagiotopoulos2000monte}%
  \BibitemOpen
  \bibfield  {author} {\bibinfo {author} {\bibfnamefont {A.~Z.}\ \bibnamefont
  {Panagiotopoulos}},\ }\href@noop {} {\bibfield  {journal} {\bibinfo
  {journal} {J. Phys.: Condens. Matter}\ }\textbf {\bibinfo {volume} {12}},\
  \bibinfo {pages} {R25} (\bibinfo {year} {2000})}\BibitemShut {NoStop}%
\bibitem [{\citenamefont {Oxtoby}(1992)}]{oxtoby1992homogeneous}%
  \BibitemOpen
  \bibfield  {author} {\bibinfo {author} {\bibfnamefont {D.~W.}\ \bibnamefont
  {Oxtoby}},\ }\href@noop {} {\bibfield  {journal} {\bibinfo  {journal} {J.
  Phys.: Condens. Matter}\ }\textbf {\bibinfo {volume} {4}},\ \bibinfo {pages}
  {7627} (\bibinfo {year} {1992})}\BibitemShut {NoStop}%
\bibitem [{\citenamefont {Hansen}\ and\ \citenamefont
  {McDonald}(2013)}]{hansen2013theory}%
  \BibitemOpen
  \bibfield  {author} {\bibinfo {author} {\bibfnamefont {J.-P.}\ \bibnamefont
  {Hansen}}\ and\ \bibinfo {author} {\bibfnamefont {I.~R.}\ \bibnamefont
  {McDonald}},\ }\href@noop {} {\emph {\bibinfo {title} {Theory of simple
  liquids: with applications to soft matter}}}\ (\bibinfo  {publisher}
  {Academic press},\ \bibinfo {year} {2013})\BibitemShut {NoStop}%
\bibitem [{Note4()}]{Note4}%
  \BibitemOpen
  \bibinfo {note} {The dependence of the growth velocity on the total
  concentration is also non-monotonic in the presence of heterogeneous
  nucleation, but the maximum velocity attained for seeded self-assembly is a
  monotonic function of $K$.}\BibitemShut {Stop}%
\bibitem [{\citenamefont {Bisker}\ and\ \citenamefont
  {England}(2018)}]{bisker2018retrieval}%
  \BibitemOpen
  \bibfield  {author} {\bibinfo {author} {\bibfnamefont {G.}~\bibnamefont
  {Bisker}}\ and\ \bibinfo {author} {\bibfnamefont {J.~L.}\ \bibnamefont
  {England}},\ }\href@noop {} {\bibfield  {journal} {\bibinfo  {journal} {Proc.
  Natl. Acad. Sci. U.S.A.}\ }\textbf {\bibinfo {volume} {115}},\ \bibinfo
  {pages} {E10531} (\bibinfo {year} {2018})}\BibitemShut {NoStop}%
\bibitem [{\citenamefont {Osat}\ and\ \citenamefont
  {Golestanian}(2022)}]{osat2022multifarious}%
  \BibitemOpen
  \bibfield  {author} {\bibinfo {author} {\bibfnamefont {S.}~\bibnamefont
  {Osat}}\ and\ \bibinfo {author} {\bibfnamefont {R.}~\bibnamefont
  {Golestanian}},\ }\href@noop {} {\bibfield  {journal} {\bibinfo  {journal}
  {Nat. Nanotech.}\ }\textbf {\bibinfo {volume} {18}},\ \bibinfo {pages} {79}
  (\bibinfo {year} {2022})}\BibitemShut {NoStop}%
\bibitem [{\citenamefont {Li}\ \emph {et~al.}(2024)\citenamefont {Li},
  \citenamefont {Xu}, \citenamefont {Zhang}, \citenamefont {Li}, \citenamefont
  {Zhang},\ and\ \citenamefont {Han}}]{li2024polymorphic}%
  \BibitemOpen
  \bibfield  {author} {\bibinfo {author} {\bibfnamefont {M.}~\bibnamefont
  {Li}}, \bibinfo {author} {\bibfnamefont {Z.}~\bibnamefont {Xu}}, \bibinfo
  {author} {\bibfnamefont {Q.}~\bibnamefont {Zhang}}, \bibinfo {author}
  {\bibfnamefont {W.}~\bibnamefont {Li}}, \bibinfo {author} {\bibfnamefont
  {Y.}~\bibnamefont {Zhang}},\ and\ \bibinfo {author} {\bibfnamefont
  {Y.}~\bibnamefont {Han}},\ }\href@noop {} {\bibfield  {journal} {\bibinfo
  {journal} {Phys. Rev. Lett.}\ }\textbf {\bibinfo {volume} {133}},\ \bibinfo
  {pages} {248202} (\bibinfo {year} {2024})}\BibitemShut {NoStop}%
\bibitem [{\citenamefont {Goodrich}\ \emph {et~al.}(2021)\citenamefont
  {Goodrich}, \citenamefont {King}, \citenamefont {Schoenholz}, \citenamefont
  {Cubuk},\ and\ \citenamefont {Brenner}}]{goodrich2021designing}%
  \BibitemOpen
  \bibfield  {author} {\bibinfo {author} {\bibfnamefont {C.~P.}\ \bibnamefont
  {Goodrich}}, \bibinfo {author} {\bibfnamefont {E.~M.}\ \bibnamefont {King}},
  \bibinfo {author} {\bibfnamefont {S.~S.}\ \bibnamefont {Schoenholz}},
  \bibinfo {author} {\bibfnamefont {E.~D.}\ \bibnamefont {Cubuk}},\ and\
  \bibinfo {author} {\bibfnamefont {M.~P.}\ \bibnamefont {Brenner}},\
  }\href@noop {} {\bibfield  {journal} {\bibinfo  {journal} {Proc. Natl. Acad.
  Sci. U.S.A.}\ }\textbf {\bibinfo {volume} {118}},\ \bibinfo {pages}
  {e2024083118} (\bibinfo {year} {2021})}\BibitemShut {NoStop}%
\bibitem [{\citenamefont {Bupathy}\ \emph {et~al.}(2022)\citenamefont
  {Bupathy}, \citenamefont {Frenkel},\ and\ \citenamefont
  {Sastry}}]{bupathy2022temperature}%
  \BibitemOpen
  \bibfield  {author} {\bibinfo {author} {\bibfnamefont {A.}~\bibnamefont
  {Bupathy}}, \bibinfo {author} {\bibfnamefont {D.}~\bibnamefont {Frenkel}},\
  and\ \bibinfo {author} {\bibfnamefont {S.}~\bibnamefont {Sastry}},\
  }\href@noop {} {\bibfield  {journal} {\bibinfo  {journal} {Proc. Natl. Acad.
  Sci. U.S.A.}\ }\textbf {\bibinfo {volume} {119}},\ \bibinfo {pages}
  {e2119315119} (\bibinfo {year} {2022})}\BibitemShut {NoStop}%
\bibitem [{\citenamefont {Chatterjee}\ and\ \citenamefont
  {Jacobs}(2025)}]{chatterjee2024multi}%
  \BibitemOpen
  \bibfield  {author} {\bibinfo {author} {\bibfnamefont {S.}~\bibnamefont
  {Chatterjee}}\ and\ \bibinfo {author} {\bibfnamefont {W.~M.}\ \bibnamefont
  {Jacobs}},\ }\href@noop {} {\bibfield  {journal} {\bibinfo  {journal} {Phys.
  Rev. X}\ ,\ \bibinfo {pages} {in press}} (\bibinfo {year}
  {2025})}\BibitemShut {NoStop}%
\bibitem [{\citenamefont {Chen}\ and\ \citenamefont
  {Jacobs}(2023)}]{chen2023programmable}%
  \BibitemOpen
  \bibfield  {author} {\bibinfo {author} {\bibfnamefont {F.}~\bibnamefont
  {Chen}}\ and\ \bibinfo {author} {\bibfnamefont {W.~M.}\ \bibnamefont
  {Jacobs}},\ }\href@noop {} {\bibfield  {journal} {\bibinfo  {journal} {J.
  Chem. Phys.}\ }\textbf {\bibinfo {volume} {158}},\ \bibinfo {pages} {214118}
  (\bibinfo {year} {2023})}\BibitemShut {NoStop}%
\bibitem [{\citenamefont {Rothemund}(2006)}]{rothemund2006folding}%
  \BibitemOpen
  \bibfield  {author} {\bibinfo {author} {\bibfnamefont {P.~W.~K.}\
  \bibnamefont {Rothemund}},\ }\href@noop {} {\bibfield  {journal} {\bibinfo
  {journal} {Nature}\ }\textbf {\bibinfo {volume} {440}},\ \bibinfo {pages}
  {297} (\bibinfo {year} {2006})}\BibitemShut {NoStop}%
\bibitem [{\citenamefont {Sun}\ \emph {et~al.}(2020)\citenamefont {Sun},
  \citenamefont {Yang}, \citenamefont {Xin}, \citenamefont {Nykypanchuk},
  \citenamefont {Liu}, \citenamefont {Zhang},\ and\ \citenamefont
  {Gang}}]{Sun2020valence}%
  \BibitemOpen
  \bibfield  {author} {\bibinfo {author} {\bibfnamefont {S.}~\bibnamefont
  {Sun}}, \bibinfo {author} {\bibfnamefont {S.}~\bibnamefont {Yang}}, \bibinfo
  {author} {\bibfnamefont {H.~L.}\ \bibnamefont {Xin}}, \bibinfo {author}
  {\bibfnamefont {D.}~\bibnamefont {Nykypanchuk}}, \bibinfo {author}
  {\bibfnamefont {M.}~\bibnamefont {Liu}}, \bibinfo {author} {\bibfnamefont
  {H.}~\bibnamefont {Zhang}},\ and\ \bibinfo {author} {\bibfnamefont
  {O.}~\bibnamefont {Gang}},\ }\href@noop {} {\bibfield  {journal} {\bibinfo
  {journal} {Nat. Commun.}\ }\textbf {\bibinfo {volume} {11}},\ \bibinfo
  {pages} {2279} (\bibinfo {year} {2020})}\BibitemShut {NoStop}%
\bibitem [{\citenamefont {Kahn}\ \emph {et~al.}(2022)\citenamefont {Kahn},
  \citenamefont {Minevich}, \citenamefont {Michelson}, \citenamefont {Emamy},
  \citenamefont {Kisslinger}, \citenamefont {Xiang}, \citenamefont {Kumar},\
  and\ \citenamefont {Gang}}]{kahn2022encoding}%
  \BibitemOpen
  \bibfield  {author} {\bibinfo {author} {\bibfnamefont {J.~S.}\ \bibnamefont
  {Kahn}}, \bibinfo {author} {\bibfnamefont {B.}~\bibnamefont {Minevich}},
  \bibinfo {author} {\bibfnamefont {A.}~\bibnamefont {Michelson}}, \bibinfo
  {author} {\bibfnamefont {H.}~\bibnamefont {Emamy}}, \bibinfo {author}
  {\bibfnamefont {K.}~\bibnamefont {Kisslinger}}, \bibinfo {author}
  {\bibfnamefont {S.}~\bibnamefont {Xiang}}, \bibinfo {author} {\bibfnamefont
  {S.~K.}\ \bibnamefont {Kumar}},\ and\ \bibinfo {author} {\bibfnamefont
  {O.}~\bibnamefont {Gang}},\ }\href@noop {} {\bibfield  {journal} {\bibinfo
  {journal} {ChemRxiv}\ } (\bibinfo {year} {2022})}\BibitemShut {NoStop}%
\bibitem [{\citenamefont {Hayakawa}\ \emph {et~al.}(2024)\citenamefont
  {Hayakawa}, \citenamefont {Videb{\ae}k}, \citenamefont {Grason},\ and\
  \citenamefont {Rogers}}]{hayakawa2024symmetry}%
  \BibitemOpen
  \bibfield  {author} {\bibinfo {author} {\bibfnamefont {D.}~\bibnamefont
  {Hayakawa}}, \bibinfo {author} {\bibfnamefont {T.~E.}\ \bibnamefont
  {Videb{\ae}k}}, \bibinfo {author} {\bibfnamefont {G.~M.}\ \bibnamefont
  {Grason}},\ and\ \bibinfo {author} {\bibfnamefont {W.~B.}\ \bibnamefont
  {Rogers}},\ }\href@noop {} {\bibfield  {journal} {\bibinfo  {journal} {ACS
  Nano}\ }\textbf {\bibinfo {volume} {18}},\ \bibinfo {pages} {19169} (\bibinfo
  {year} {2024})}\BibitemShut {NoStop}%
\bibitem [{\citenamefont {Videb{\ae}k}\ \emph {et~al.}(2024)\citenamefont
  {Videb{\ae}k}, \citenamefont {Hayakawa}, \citenamefont {Grason},
  \citenamefont {Hagan}, \citenamefont {Fraden},\ and\ \citenamefont
  {Rogers}}]{videbaek2024economical}%
  \BibitemOpen
  \bibfield  {author} {\bibinfo {author} {\bibfnamefont {T.~E.}\ \bibnamefont
  {Videb{\ae}k}}, \bibinfo {author} {\bibfnamefont {D.}~\bibnamefont
  {Hayakawa}}, \bibinfo {author} {\bibfnamefont {G.~M.}\ \bibnamefont
  {Grason}}, \bibinfo {author} {\bibfnamefont {M.~F.}\ \bibnamefont {Hagan}},
  \bibinfo {author} {\bibfnamefont {S.}~\bibnamefont {Fraden}},\ and\ \bibinfo
  {author} {\bibfnamefont {W.~B.}\ \bibnamefont {Rogers}},\ }\href@noop {}
  {\bibfield  {journal} {\bibinfo  {journal} {Sci. Adv.}\ }\textbf {\bibinfo
  {volume} {10}},\ \bibinfo {pages} {eado5979} (\bibinfo {year}
  {2024})}\BibitemShut {NoStop}%
\bibitem [{\citenamefont {Duque}\ \emph {et~al.}(2024)\citenamefont {Duque},
  \citenamefont {Hall}, \citenamefont {Tyukodi}, \citenamefont {Hagan},
  \citenamefont {Santangelo},\ and\ \citenamefont {Grason}}]{duque2024limits}%
  \BibitemOpen
  \bibfield  {author} {\bibinfo {author} {\bibfnamefont {C.~M.}\ \bibnamefont
  {Duque}}, \bibinfo {author} {\bibfnamefont {D.~M.}\ \bibnamefont {Hall}},
  \bibinfo {author} {\bibfnamefont {B.}~\bibnamefont {Tyukodi}}, \bibinfo
  {author} {\bibfnamefont {M.~F.}\ \bibnamefont {Hagan}}, \bibinfo {author}
  {\bibfnamefont {C.~D.}\ \bibnamefont {Santangelo}},\ and\ \bibinfo {author}
  {\bibfnamefont {G.~M.}\ \bibnamefont {Grason}},\ }\href@noop {} {\bibfield
  {journal} {\bibinfo  {journal} {Proc. Natl. Acad. Sci. U.S.A.}\ }\textbf
  {\bibinfo {volume} {121}},\ \bibinfo {pages} {e2315648121} (\bibinfo {year}
  {2024})}\BibitemShut {NoStop}%
\bibitem [{\citenamefont {Kuznetsov}(1998)}]{Kuznetsov1998}%
  \BibitemOpen
  \bibfield  {author} {\bibinfo {author} {\bibfnamefont {Y.~A.}\ \bibnamefont
  {Kuznetsov}},\ }\href@noop {} {\emph {\bibinfo {title} {Elements of applied
  bifurcation theory}}},\ \bibinfo {edition} {2nd}\ ed.,\ Applied Mathematical
  Sciences\ (\bibinfo  {publisher} {Springer},\ \bibinfo {address} {New York,
  NY},\ \bibinfo {year} {1998})\ pp.\ \bibinfo {pages} {83--86}\BibitemShut
  {NoStop}%
\end{thebibliography}
\end{document}


\maketitle
\onecolumngrid
\renewcommand\thefigure{S\arabic{figure}}    
\setcounter{figure}{0}
\renewcommand\theequation{S\arabic{equation}}    
\setcounter{equation}{0}
\renewcommand\thetable{S\arabic{table}}
\setcounter{table}{0}

\section*{Supplementary Information for ``Dynamical phase transition in the growth of programmable polymorphic materials''}

\section{Calculating the storage capacity in the dynamical mean-field model}

\begin{figure}
    \centering
    \includegraphics[width=.75\columnwidth]{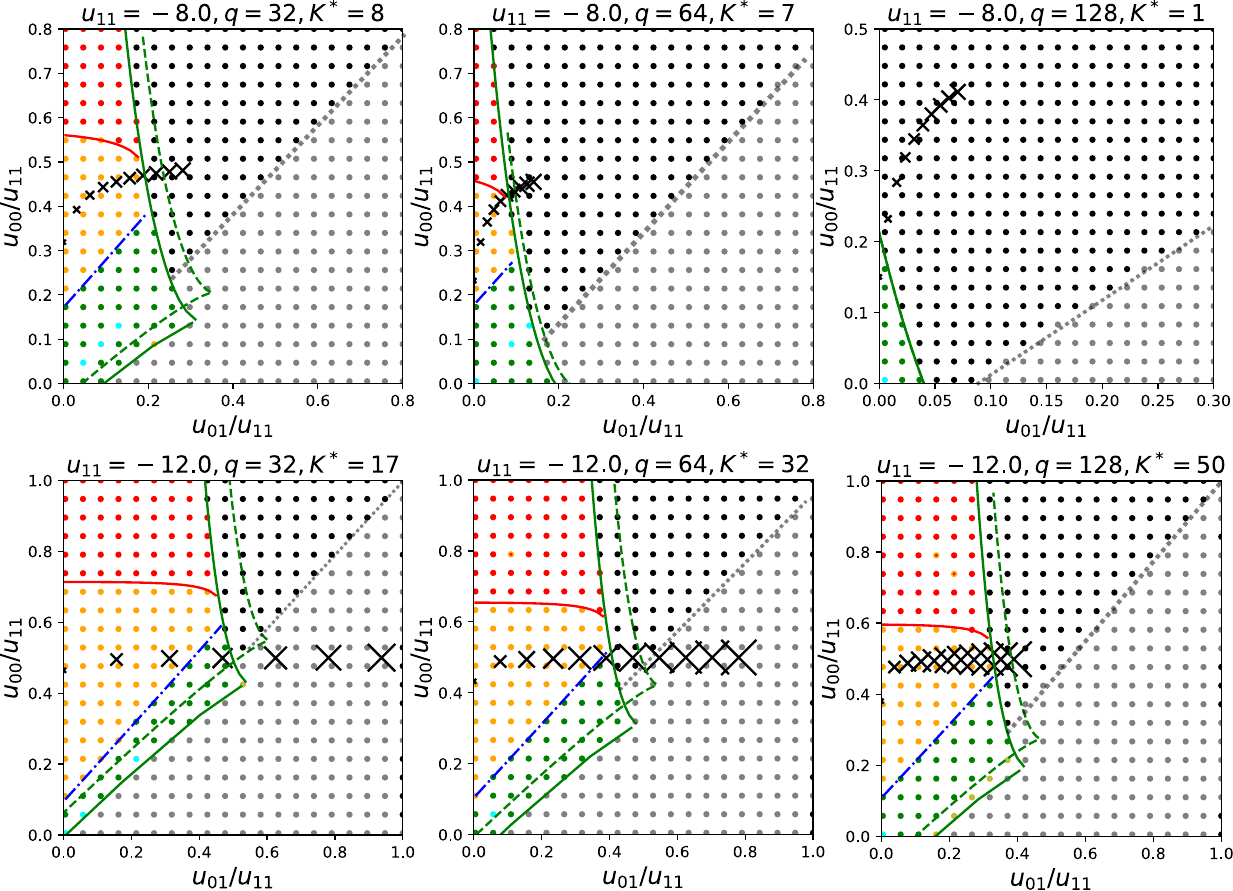}
    \caption{\textbf{Dynamical phase diagrams with different bond strengths and numbers of components.}
      The top and bottom rows correspond to ordered interaction parameters $u_{11} = -8$ and $u_{11} = -12$, respectively.
      From left to right, the number of components $n$ increases, where $q = zn$ with $z = 4$.
      The solid green line indicates the boundary of the region with two stable coexistence points when $K = 1$, whereas the dashed green line indicates this boundary when $K$ is set equal to the storage capacity, $K = K^*$.
      The points are colored using the same convention as in Fig.~1B in the main text, except that cyan points denote instances where the numerical solver did not converge.
      The crosses indicate the $K$-dependent interactions, starting with $K = 1$ with increments of one (i.e., $K = 1, 2, 3, \ldots$) and five (i.e., $K = 1, 6, 11, \ldots$) in the top and bottom rows, respectively.}
    \label{sfig:scan}
\end{figure}

The topology of the dynamical phase diagram shown in Fig.~1B in the main text remains the same when the bond strength, $\epsilon$, and the number of components, $n$, change (\figref{sfig:scan}). 
In general, increasing the number of components shifts the ordered assembly region in the direction of weaker disorder--disorder interactions, whereas increasing the bond strength shifts the ordered assembly region in the direction of stronger cross-talk interactions.

\begin{figure}
    \centering
    \includegraphics[width=.5\columnwidth]{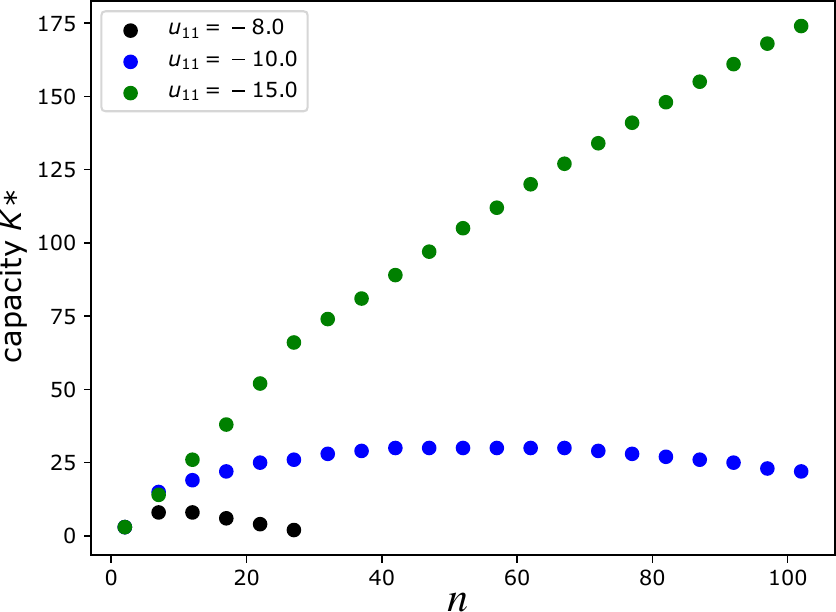}
    \caption{\textbf{Scaling of the storage capacity.}
      As the order--order interaction strength increases, the scaling of the storage capacity, $K^*$, approaches a linear function of the number of components, $n$.  Points indicate numerical solutions, accounting for the $K$ dependence of both the mean-field interaction parameters and the dynamical mean-field equations, as described in the text.  The capacity can be a nonmonotonic function of $n$ for small $|u_{11}|$ due to the nonlinearity of the disorder--disorder interaction parameter, Eq.~(D2) in the main text, in this regime.}
    \label{sfig:scaling}
\end{figure}

The storage capacity, $K^*$, is reached when the number of coexistence points goes from two to one, signaling that positive growth velocities are not possible for the ordered structures.
[In the analysis that follows, we do not distinguish whether $c_{\text{coex}}^{\text{disord}}$ is less than or greater than $c_{\text{coex}}^{\text{ord}}$, since the boundary between the dynamical transition region and the disordered assembly region is not crossed by the parametrized curve $(u_{01}(K), u_{00}(K))$ in the range of bond strengths and component counts that we consider.]
We therefore determine the storage capacity by numerically finding the largest value of $K$ for which the parametrized curve $(u_{01}(K), u_{00}(K))$ yields two coexistence points.
This calculation uses the $K$-dependent form of the dynamical mean-field equations (see App.~A in the main text), to ensure that we account for all contributions due to changing $K$.
Numerically determined values of $K^*$ are shown for example dynamical phase diagrams in \figref{sfig:scan}.
Interestingly, $K^*$ can be a non-monotonic function of the number of components, $n$, when the order--order interaction strength is relatively weak, as shown in \figref{sfig:scaling}.

Nonetheless, in the strong bond limit, i.e., $\epsilon \rightarrow -\infty$, we predict that the storage capacity scales linearly with the number of components, $n$.
This result is due to the assumption in our model that every bond corresponds to a unique polymorph.
Specifically, in the strong-bonding limit, the parametrization of the mean-field interactions leads to $u_{00} \to \nicefrac{1}{2}$ for disordered interactions at the interface of a two-dimensional structure.
The storage capacity in this limit is therefore determined by the interplay between the cross-talk interaction, $u_{01}$, and the boundary of the region where there are two stable coexistence points.
Suppose that for a fixed bond strength $\epsilon$, a system with $n = q/z$ components has a storage capacity of $K^*$, which corresponds to a specific point in the $u_{01}/u_{11} - u_{00}/u_{11}$ plane where the $u_{01}$ parametrization intersects the boundary of the region with two stable coexistence points.
We then consider changing $n \to n + \delta n$, where $\delta n / n$ is small.
This implies $q \to q + \delta q$, so that to first order $u_{01} \to u_{01} [1 + \delta q / (q-1)]$.
Thus, the perturbation $q \to q + \delta q$ results in stronger cross-talk interactions.
Next, the boundary of the region with two stable coexistence points depends on $b$ through the $r$-Lambert function, which to first order is linear in its argument, $-e^{-b+1}$.
In particular, in the limit of large $q$, we expand $-e^{-b} \sim -1 + b$ to obtain $-e^{-b} \to b + \delta q[1 / (q-K) + u_{01} / (q-1)] -1$, since $b \to b + \log[1 + \delta q / (q-K)] + \delta u_{01} \sim b + \delta q[1 / (q-K) + u_{01} / (q-1)]$.
Transforming from the $a-b$ plane back to the $u_{01}/u_{11} - u_{00}/u_{11}$ plane, this perturbation implies that the boundary of the region with two stable coexistence points shifts in the direction of lower disorder--disorder and cross-talk interactions.
Because both perturbations are linear with respect to $\delta q/q$, we predict an asymptotically linear scaling of the storage capacity, $K^*$, in the limit of large $q$ and strong-bonding interactions.

\section{Accounting for diffusion at the solid--dilute interface in the dynamical mean-field model}

We now consider the effect of surface diffusion at the solid--dilute interface.
In the model presented in the main text, the net attachment rate to the coarse-grained state $\alpha$ from the dilute phase is given by $\Gamma_\alpha$, where monomer diffusion in the dilute phase sets the absolute time units.
More specifically, the flux $\Gamma_\alpha$ has units of $\tau_D^{-1} \simeq 4D / \lambda^2$, where $D$ is the monomer self-diffusion coefficient in the dilute phase and $\lambda \sim c^{-1/2}$ is the typical distance between monomers in the dilute phase.
To include surface diffusion in the model, we consider a scenario in which an $\alpha$-state subunit detaches from the bulk structure, remains nonspecifically adsorbed to the interface, and subsequently rearranges to align with the $\beta$ coarse-grained state.
 We denote the ratio of the time units for this surface rearrangement relative to $\tau_D^{-1}$ by $k_{\alpha\beta}$. 
The governing equation for the dynamical mean-field model then becomes
\begin{equation}
\dot{p_\alpha} = \Gamma_\alpha +\sum_{\beta=0}^K k_{\beta \alpha} p_\beta e^{\sum_{l=0}^K u_{l\beta} p_l} - p_\alpha  e^{\sum_{l=0}^K u_{l\alpha} p_l}   \sum_{\beta=0}^K k_{\alpha \beta} - p_\alpha \dot{N},
\end{equation}
where the second and third terms account for transitions to and from, respectively, the $\alpha$ coarse-grained state via surface diffusion.
For the $K = 1$ case, this reduces to 
\begin{subequations}
\begin{align}
\dot{p} &= \Gamma_1 +  k_{1 1} p e^{u_{11}p + u_{01}(1-p)} + k_{0 1} (1-p) e^{u_{00} (1-p) + u_{01} p}  - ( k_{1 1} +  k_{1 0}) p e^{u_{11}p  + u_{01}(1-p)}  - p \dot{N} \label{eq:diffusion} \\
&= \Gamma_1  - p \dot{N} +  k_{0  1} (1 - p) e^{(u_{01} - u_{00})p + u_{00}}  -  k_{10} p e^{(u_{11} - u_{01})p + u_{01}}.
\end{align}
\end{subequations}
At coexistence, the solution has to satisfy $\dot{p} = 0, \dot{N} = 0$.
Moreover, coexistence solutions from the original model must also be solutions to the model with surface diffusion, since detailed balance is satisfied at coexistence and thus the incorporation of additional kinetic pathways does not affect the equilibrium behavior.
Thus, by applying $\dot{p} = 0$ and $\Gamma_1 = 0$ at coexistence, we find that $k_{10}/k_{01} = q-K$.
With surface diffusion, the equation for the bifurcation point, $\partial g / \partial p = 0$, becomes
\begin{align}
\frac{\partial g}{\partial p} = 0 = &-c + (u_{01} - u_{00}) e^{(u_{01} - u_{00})p + u_{00}} k_{01} \label{eq:dynamical_ptb} \\
& + e^{(u_{11} - u_{01})p + u_{01}} \left[ (2p - 1) - p(u_{11} - u_{01}) (1 + k_{10}) + p^2 (u_{11} - u_{01}) - k_{10}\right] \nonumber  \\
& - e^{(u_{01} - u_{00})p + u_{00}} \left[ 
(2p - 1) - p(u_{01} - u_{00}) (1 - k_{01}) + p^2 (u_{01} - u_{00}) + k_{01}
\right]. \nonumber
\end{align}
This equation has the same structure as the equation without surface diffusion and reduces to Eq.~(A9b) in the main text in the limit $k_{01} \to 0$.
Thus, in parameter regimes where bifurcation occurs, the bifurcation diagram is qualitatively the same as that of the system without surface diffusion, sharing precisely the same coexistence points and a perturbed bifurcation point.
As a result, the only quantitative change to the dynamical phase diagram is the boundary between the ordered assembly and the dynamical phase transition regions, given by Eq.~(B3) in the main text, which is minimally altered by the perturbed bifurcation point concentration when $k_{01} \ll 1$.

\section{Lattice model implementation and simulation details}

\subsection{Directional interactions and unit cell designs}
To implement the lattice model with sticker-specific directional interactions, we use pseudo-species to represent the components with different orientations.
There are $q + 1$ possible states for a lattice site, where $q=zn$ with state $0$ corresponding to a vacancy and states $1, \ldots, zn$ corresponding to occupied states. 
For the two-dimensional square lattice, the coordination number is $z=4$, and the state $4(\alpha-1)+r+1$ corresponds to the component type $\alpha = 1, \ldots, N$ and the orientational state $r = 0, 1, 2, 3$. 
The orientational state determines the position of sticker 0 (i.e., N in Fig.\ 3A in the main text).
The remaining stickers $1$ (E), $2$ (S), and $3$ (W) follow in clockwise order.

A unit cell design can be specified by the relative positions and orientations of all components within the unit cell, where nearest-neighbor (including periodic image nearest-neighbor) stickers bind.
Each unit cell design corresponds to an interaction tensor $U^{(\alpha)}$, where $\alpha = 1, \ldots, K$ denotes the unit cell (i.e., crystal polymorph) index.
In practice, the interaction tensor is represented by a $q \times q$ matrix, which is indexed according to the sticker labels on the tiles defined above.
In this study, we design a set of $K$ unit cells that have no shared bonds.
The corresponding unit cell type can thus be uniquely identified for every bond.
For simplicity, we assume that all bonds have the same energy $\epsilon$.
All interactions with vacant lattice sites are set to zero.
The total interaction tensor $U$ is the union of the $K$ interaction tensors for the encoded designs, $\{U^{(\alpha)}\}$.

We perform simulations in the grand canonical ensemble.
The Hamiltonian of the lattice model is
\begin{equation}
  H = \sum_{<l, l^\prime>}  U_{\text{species}(l)_{\text{sticker}(l)}, \text{species}(l^\prime)_{\text{sticker}(l^\prime)}} S_l S_{l^\prime} + \mu \sum_{l}  S_{l^\prime},
\end{equation}
where the indicator function $S_{l} = 1$ if a lattice site $l$ is occupied or $0$ if it is vacant.
The first sum runs over all nearest-neighbor lattice-site pairs, taking into account periodic boundary conditions for the transverse dimension of the elongated lattice.
When both sites are occupied, the species types and orientations, together with the relative positions of sites $l$ and $l^\prime$, are used to determine the sticker labels facing one another between lattice sites $l$ and $l^\prime$.  
The second sum runs over all lattice sites.
We assume that all components have the same chemical potential, $\mu$. 

\begin{figure}
  \includegraphics[width=\columnwidth]{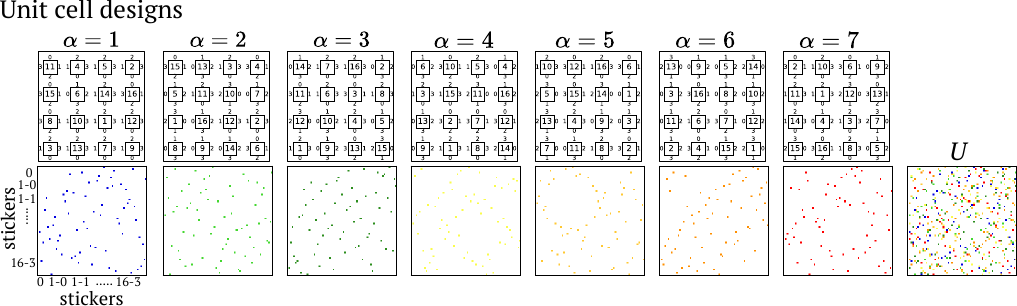}
  \caption{\textbf{Multicomponent unit cell designs and interaction tensors.}
    \textit{Top row:} The seven unit cell designs used in this study.  Tiles are labeled by their component indices.  The tile orientations are indicated by the positions of the stickers on the sides of the tiles, were $0$ represents N.
    \textit{Bottom row:} The interaction tensor corresponding to each unit cell above, shown in matrix form in terms of the pseudo-species (either 0 or $\alpha$-$r$), as defined in the text.  The total interaction tensor, $U$, is shown in the final column.  Bonds are colored in accordance with the simulation snapshots shown in the main text.
    \label{sfig:designs}}
\end{figure}

In addition to the no-shared-bond constraint, we also seek to minimize differences among the equilibrium stabilities of the designed polymorphs.
To this end, we choose unit cell designs that do not contain ``fully coordinated point defects'', meaning that a tile within a designed unit cell can either be an incorrect component or have an incorrect orientation and still form four bonds.
Generating designs that satisfy this constraint requires that the $K$ unit cells are designed as a set.
We therefore perform simulated annealing~\cite{kirkpatrick1983sa} in order to find designs with zero shared bonds and zero fully coordinated point defects by simultaneously permuting and rotating the tiles in all $K$ unit cells.
The complete set of seven unit cell designs used for generating all the simulation data in this study are shown in \figref{sfig:designs}.

\subsection{Kinetic Monte Carlo (kMC) simulations}
We utilize the kinetic Monte Carlo (kMC) method~\cite{gillespie1977exact} to simulate stochastic dynamics under the assumption that changes to the system occur with known transition rates between discrete states.
Given the current state of the system, the next reaction will occur with probability proportional to its rate within the time interval $[t, t + \Delta t)$, where $\Delta t$ is exponentially distributed with respect to the total rate of all possible reactions.
The possible reactions in our model are the insertion of a tile (of any species type and any orientation into any empty lattice site) with the rate $k_\text{attach} = \exp(\mu / k_{\text{B}}T)$, and the removal of a tile (from any occupied lattice site) with the rate $k_\text{detach} = \exp(n_\text{b} \epsilon / k_{\text{B}}T)$, where $n_\text{b}$ is the number of nearest-neighbor bonds that the tile forms.
The bond strength is chosen to be $\epsilon = -4 k_{\text{B}}T$ throughout all our simulations.
We set the fundamental time unit equal to that of the slowest possible reaction, $\tau_0 = 1/\exp(3 \epsilon / k_{\text{B}}T)$.
All times reported throughout the paper are given in units of $\tau_0$.

\subsection{Coexistence simulations for locating the dilute--solid coexistence curve}
\begin{figure}
  \includegraphics[width=\columnwidth]{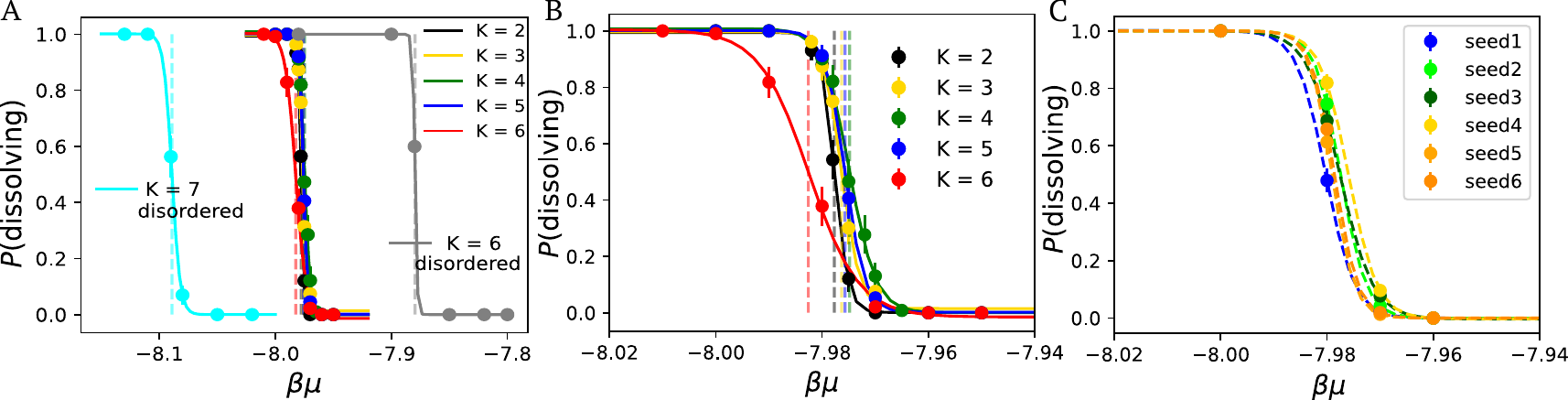}
  \caption{\textbf{Determination of coexistence chemical potentials for designed polymorphs and disordered structures.} 
  (A)~The commitment probability, $P(\text{dissolving})$, for dissolution of the initial structure using $K = 2, \ldots, 7$ unit cell designs (see text). The initial ordered crystal is the type-1 unit cell design for $K = 2, \ldots, 6$. 
  (B)~A zoomed-in view of the commitment probability for dilute--ordered coexistence.
  (C)~The commitment probability for structure dissolution with $K = 6$ encoded unit cell designs, where each unit cell design, $\alpha = 1, \ldots, 6$, is used as the initial condition for the calculation.
    Here, $\beta = 1/k_{\text{B}}T$ is the inverse temperature.
    Error bars are calculated by bootstrapping the data five times and taking the standard deviation. 
    \label{sfig:mu-coex}} 
\end{figure}

We perform direct-coexistence~\cite{panagiotopoulos2000monte} kMC simulations to map out the equilibrium coexistence curve between the dilute phase and either an ordered crystal or the disordered structure.
We initialize a $40 \times 40$ lattice with periodic boundary conditions that is half empty and half filled with either one of the designed crystal polymorphs or the disordered structure.
We calculate the commitment probability that the seed structure dissolves, leaving the entire lattice in the dilute phase, by performing 100 independent simulations.
We then fit the commitment probability to a sigmoid function to extract $\mu_{\text{coex}}$, at which point the probability of dissolution is 50\% (\figref{sfig:mu-coex}).
For the disordered structure calculations, the initial configurations for the filled half of the lattice are obtained by equilibrating the $K = 7$ mixture at $\mu = -8 k_{\text{B}}T$.

\subsection{Nucleation simulations for locating the constant-rate homogeneous nucleation line}
\begin{figure}
  \includegraphics[width=0.5\columnwidth]{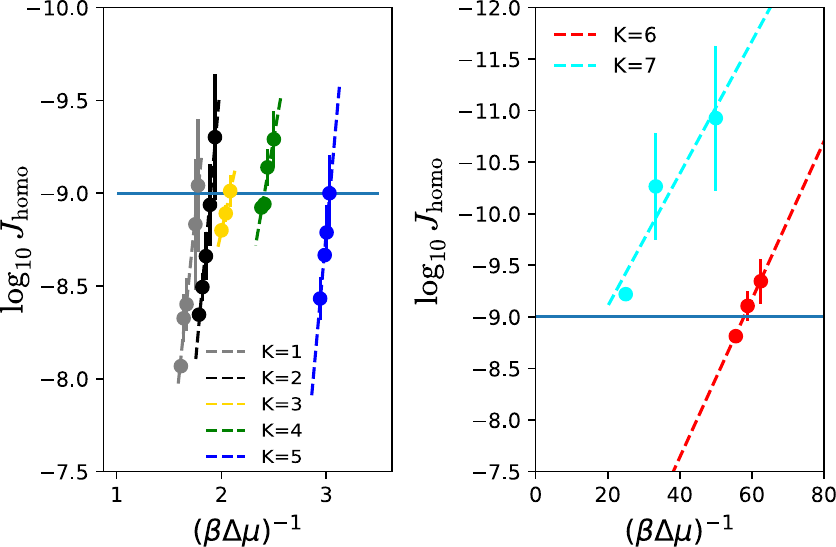}
  \caption{\textbf{Determination of the constant-rate homogeneous nucleation boundary.}
    The logarithm of the nucleation rate, $J_{\text{homo}}$, is plotted relative to $\Delta\mu / k_{\text{B}}T = \log (c / c_{\text{coex}})$ for each value of $K$.  According to classical nucleation theory, the slope of each line is proportional to the surface tension, which is roughly constant for $K \le 5$.  The $K=7$ results correspond to nucleation of the disordered structure.
    Here, $\beta = 1/k_{\text{B}}T$ is the inverse temperature.
    Error bars are calculated by bootstrapping the data 20 times and taking the standard deviation.
    The slope of logarithm nucleation rate with respect to the inverse of the excess chemical potential is proportional to the surface tension between the gas phase and the nucleus. 
    The ratio of the slope for the $K = 6$ and $K = 7$ cases to that for the $K = 1, \ldots, 5$ cases is of order 1/100.
    Consistent with this difference, the nuclei observed in the former cases ($K = 6$ and $K = 7$) consist of mostly disordered structures, whereas the nuclei in the latter cases are mostly single polymorphs.
    We therefore infer that the disordered structures have substantially lower surface tension than the ordered polymorphs.
  \label{sfig:homo-nuc}} 
\end{figure}

To estimate the homogeneous nucleation rate, we perform kMC simulations starting with an empty box with periodic boundary conditions and dimensions $56 \times 100$, which is half the size of the moving window utilized in the growth simulations described below. 
We measure the size of the largest cluster in this system every $10^7$ time units, and we register a nucleation event if we find that this cluster size reaches $25\%$ of the simulation box volume.
Assuming that classical nucleation theory holds~\cite{oxtoby1992homogeneous}, such that the nucleation rate varies exponentially with respect to the reciprocal of the chemical potential difference in a two-dimensional system, we fit the nucleation rate and interpolate to find the chemical potential at which $\log_{10} J_\text{homo} = -9$ (\figref{sfig:homo-nuc}).
This procedure is repeated for each value of $K$ to identify the line of constant homogeneous nucleation rate in Fig.\ 4 in the main text.

\subsection{Growth simulations}
To simulate crystal growth, we perform kMC in a simulation box with periodic transverse dimension $L_y = 56$ and open horizontal dimension $L_x = 1000$.
The growth process is initialized with a selected seed crystal filled up to column 160, and the simulation is terminated when the front of the bulk structure reaches $80\%$ of the horizontal dimension.
We impose a kinetic constraint in which fully-coordinated lattice sites are ``frozen''~\cite{whitelam2014critical}, in which case no transitions can take place.
This kinetic constraint satisfies detailed balance and tends to inhibit equilibration deep within the bulk solid.

To speed up the simulations, we define a moving window centered at the bulk-structure growth front, defined as the interface between the growing bulk structure and the dilute phase.
This interface can be identified by performing a connected component search~\cite{tarjan1971dfs} starting from the dilute phase.
In practice, we only allow transitions at lattice sites within the moving window.
This choice allows us to explicitly simulate the dilute phase and the growing structure near the interface, while ignoring the extremely slow equilibration of the bulk crystal behind the moving window and the diffusion of the tiles in the dilute phase far in front of the moving window.
To further reduce computational cost, we update the window position at a preset frequency.
The window width and the update frequency are chosen such that the distribution of interfacial widths (defined as the difference between the maximum and minimum horizontal positions of the instantaneous bulk--dilute interface) is statistically indistinguishable from simulations performed without the moving window.
For all data shown, the window width is chosen to be 200 lattice sites and the window is updated every 1000 kMC moves.

In the data presented in the main text, the initial crystal seed is chosen to be the $\alpha = 1$ polymorph. 
Seeding with different crystal polymorphs can adjust the coexistence curve and the constant-rate nucleation boundaries slightly due to minor differences in the thermodynamic stabilities of the crystal polymorphs.
Nonetheless, the qualitative picture presented in the main text is consistent regardless of the initial seed polymorph.

\subsection{Simulations of heterogeneous nucleation at the bulk interface}

To obtain the heterogeneous nucleation rate (i.e., the rate at which the ordered crystal that spans the transverse dimension switches between polymorphs) from the growth trajectory, we first identify the bulk structure by performing a connected cluster search starting from the seed crystal.
We then find all clusters of lattice sites that correspond to distinct polymorphs and check whether any cluster that differs from the seed polymorph spans the transverse dimension. 
If such a cluster exists, we identify this point in the growth trajectory as a switching event, record the time, and terminate the simulation.

\begin{figure}
  \includegraphics[width=0.55\columnwidth]{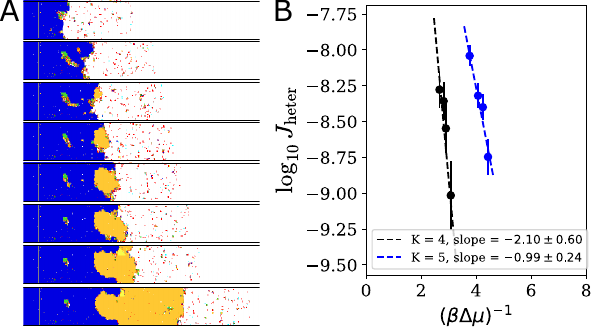}
  \caption{\textbf{Heterogeneous nucleation at the crystal--dilute interface during ordered crystal growth.}
    (A)~An example $K = 5$ trajectory showing the stochastic nucleation of a non-seed-type nucleus (yellow) at the growth front of the seed polymorph (blue).
    (B)~Maximum likelihood estimate of the heterogeneous nucleation ``switching rate'' and the fit to the two-dimensional classical nucleation theory functional form, where $\Delta\mu / k_{\text{B}}T = \log (c / c_{\text{coex}})$ (dashed lines).
    Here, $\beta = 1/k_{\text{B}}T$ is the inverse temperature.
    Error bars are calculated by bootstrapping the data 20 times and taking the standard deviation. 
    \label{sfig:heternuc}}
\end{figure}

Assuming that the switching events are Poisson distributed, the maximum likelihood estimate (MLE) of the switching rate, $(\tau_{\text{MLE}})^{-1}$, is given by 
\begin{equation}
  \tau_\text{MLE} = \begin{cases}
    \frac{1}{n_\text{succ}}(\sum_k \tau_{\text{succ}, k} + \sum_{k^\prime}\tau_{\text{fail}, k^\prime}), & \text{if}\,\, {n_\text{succ} > 0}, \\
    \sum_{k^\prime}\tau_{\text{fail}, k^\prime} & \text{otherwise},
  \end{cases}
\end{equation} 
where $n_\text{succ}$ and $n_\text{fail}$ are the numbers of trials containing or not containing switching events, respectively.
Similarly, $\tau_\text{succ}$ and $\tau_\text{fail}$ are the times at which the switching event happens or the trajectory ends without a switching event, respectively.
An example trajectory showing a heterogeneous nucleation event is illustrated in \figref{sfig:heternuc}A.
As in our analysis of homogeneous nucleation, we find the line of constant-rate heterogeneous nucleation by fitting to the prediction of classical nucleation theory and finding the chemical potential at which $\log_{10} J_\text{heter} = -9$ (\figref{sfig:heternuc}B).

\subsection{Simulations of the dynamical first-order phase transition}

To calculate average growth rates and interfacial properties at the storage capacity, $K^* = 6$, we first locate the bulk--dilute interface via a connected component search starting from the dilute phase.
The rightmost column of this interface is identified as the position of the growth front.
The overall growth rate of the bulk structure is then obtained via a least-squares fit of the growth front versus time.
To demonstrate the possibility of dynamical coexistence, we start the growth trajectory from either the ordered seed configuration or from an initial disordered configuration obtained by running the growth simulation at $\mu = -7.89 k_{\text{B}}T$, where we first observe time-extensive disordered growth.

To determine the disordered layer width, we first identify whether occupied lattice sites are in an ordered or disordered state based on their local environments.
Here the local environment is defined to be the 5 x 5 grid of lattice sites centered at the lattice site of interest.
The front-most boundary of the ordered crystal is located by searching for the connected ordered structure boundary starting from the initial seed and selecting the rightmost column for every row.
The front-most boundary of the disordered structure is similarly located by searching for the connected disordered structures starting from the dilute phase and selecting the rightmost column for every row.
The width of the disordered wetting layer is then defined to be the average difference between the locations of the ordered and disordered boundaries over the rows containing the disordered structure.

Skipping the initial $10^8$ kMC time for each trajectory to ensure that the growth process has reached steady state, we report the average growth rates and layer widths obtained from five independent simulations for each concentration in Fig.\ 5A,B in the main text. 
Each simulation in the vicinity of the dynamical transition ran for two weeks on a single CPU to collect sufficient statistics.
The error bars indicate the standard deviation of five independent runs.

\begin{figure}
  \includegraphics[width=\columnwidth]{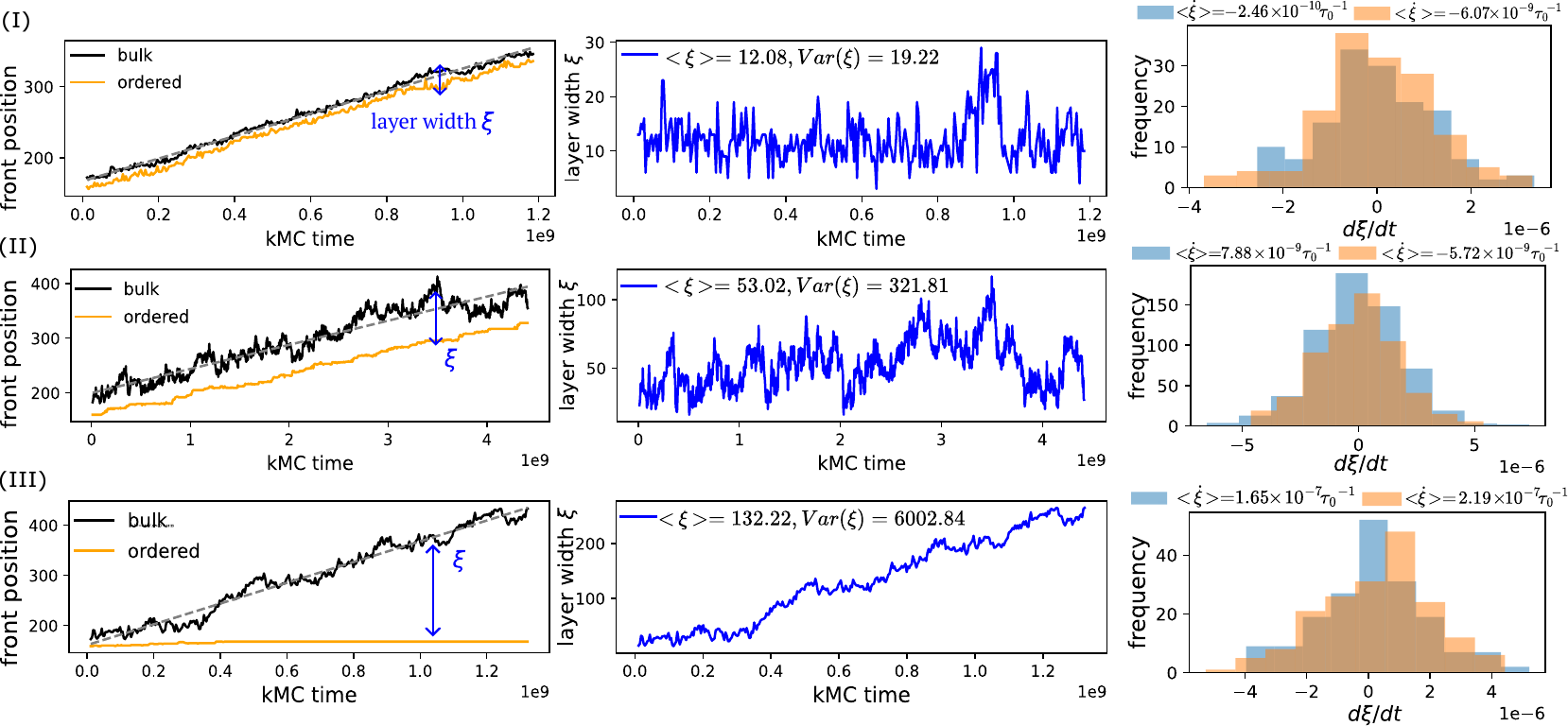}
  \caption{\textbf{Time series analyses of the growth front and interface near the dynamical first-order phase transition.}
  The top, middle, and bottom rows correspond to the chemical potentials $\mu / k_{\text{B}}T = \log c = -7.90$, $-7.892$, and $-7.89$, respectively, for the $K=6$ system.
  \textit{Left:}~Time series data for the bulk-structure (black) and ordered (orange) growth fronts.
  \textit{Center:}~Time series data for the difference between the bulk and ordered growth front positions, corresponding to the disordered wetting layer width, $\xi$.
  \textit{Right:}~Distributions showing the instantaneous rate of change of the interfacial width, $d\xi/dt$, obtained from the first (blue) and second (orange) halves of the time series data shown to the left.  These distributions are statistically indistinguishable, as should be expected for steady-state growth.
  \label{sfig:traj}}
\end{figure}

At steady state, the bulk growth rate and the average disordered wetting layer width should both be constant in time.  
In the cases of stable seeded self-assembly and disordered structure growth, it is easy to verify that these conditions are met (\figref{sfig:traj}).
However, as it takes significantly longer to simulate self-assembly in the vicinity of the dynamical phase transition, and because the fluctuations of the wetting layer tend to be large, greater care is required to confirm that the growth process has reached steady state.
We therefore perform stationary tests on the time series of the wetting layer width using the Augmented Dickey--Fuller test (ADF Test)~\cite{dickey1979ADF} and the Kwiatkowski--Phillips--Schmidt--Shin test (KPSS test)~\cite{kwiatkowski1992KPSS}, which are standard statistical tests used in time series analyses.
The stationary tests are based on an autoregressive model of the time series data relative to a time-shifted version, and the lags used correspond to the number of shifting steps applied to the original data.
Here the wetting layer width is defined by the difference between the rightmost column of the disordered front boundary and the ordered front boundary for simplicity, to avoid the need to average over the transverse dimension. 
The ADF statistic tests the null hypothesis that the time series has a unit root against the alternative of no unit root. 
The presence of a unit root implies that the variance depends on time and that the time series is therefore not stationary.
For example, a random walk is a unit root process with a constant mean but a diverging variance as time increases.
The KPSS statistic tests the null hypothesis that the time series is stationary around a deterministic trend against the alternative that the series has a unit root.
The null hypotheses are rejected when the test statistic is more negative (greater) than the critical value in the ADF (KPSS) test, with a p-value less than a significance level of 0.05.

The results shown in Table~\ref{tab:test1} indicate that the ADF test statistic is more negative than the critical value and that the p-value is less than 0.05.
We therefore reject the null hypothesis for the ADF test.
Meanwhile, the results shown in Table~\ref{tab:test2} indicate that the KPSS test statistic is greater than the critical value and that the p-value is less than 0.05.
We therefore also reject the null hypothesis for the KPSS test.
Combining these two tests, we conclude that the trajectories may have unit roots but do not have a deterministic trend, suggesting that the layer width is undergoing a random walk at steady state without a bias towards growing or shrinking over time.
These findings support our conclusion that the disordered layer is a stable wetting layer for growth at concentrations below $c_{\text{trs}}$ in Fig.\ 5C of the main text.

\begin{table}[!htb]
    \begin{minipage}{.5\linewidth}\centering
        \begin{tabular}{|c|c|c|c|} 
        \hline
             & Trajectory~1 & Trajectory~2 & Trajectory~3 \\
             \hline
            Critical Value ($5\%$) & -2.864  & -2.864  & -2.866  \\
            \hline
            Test Statistic & -4.635 & -3.742 & -3.041 \\
            \hline
            p-value & 0.00011 &  0.0035 & 0.0313 \\
            \hline
            Lags Used & 2.0 & 2.0 & 2.0\\
            \hline
        \end{tabular}
        \caption{ADF test applied to time series from three independent trajectories at $\mu = -7.892 / k_{\text{B}}T$.}
        \label{tab:test1}
    \end{minipage}

    \vskip2ex
    
    \begin{minipage}{0.5\linewidth}\centering
        \begin{tabular}{|c|c|c|c|}
        \hline
             & Trajectory~1 & Trajectory~2 & Trajectory~3 \\
             \hline
            Critical Value ($5\%$) & 0.463 & 0.463 & 0.463 \\
             \hline
            Test Statistic & 1.462 &  0.611 & 0.777\\
            \hline
            p-value & 0.010 & 0.022 & 0.010\\
            \hline
            Lags Used & 20.0 & 20.0 & 20.0\\
            \hline
        \end{tabular}
        \caption{KPSS test applied to time series from three independent trajectories at $\mu = -7.892 / k_{\text{B}}T$.}
        \label{tab:test2}
    \end{minipage} 
\end{table}